\documentclass[12pt,a4paper,final]{iopart}

\usepackage{amsmath}
\usepackage{esint} 

\pdfoutput=1
\def\contentsname{\empty}
\usepackage{graphicx}
\usepackage{cite}

\usepackage{graphicx}
\usepackage{color}
\usepackage{hyperref}
\usepackage{url}
\usepackage{appendix}

\newcommand{\ea}[1]{\begin{align}#1\end{align}}

\newcommand{\nn}{\nonumber}

\newcommand{\phset}{\mathbf{p},\mathbf{h} }
\newcommand{\smallk}{\xrightarrow{p\rightarrow h}}

\newcommand{\resatilde}{\tilde{a}_{\text{res}(h_i)}}

\newcommand{\pint}{{\rm P}\int}
\newcommand{\ppint}{{\rm P}_+\int}

\allowdisplaybreaks

\begin{document}

\title[Thermodynamic form factors in the Lieb-Liniger model]{Particle-hole pairs and density-density correlations in the Lieb-Liniger model}

\author{J. De Nardis$^1$ and M. Panfil$^2$}

\address{$^1$D\'epartement de Physique, Ecole Normale Sup\'erieure, PSL Research University, CNRS, \\ 24 rue Lhomond, 75005 Paris, France}

\address{$^2$Institute of Theoretical Physics, University of Warsaw,\\
 ul. Pasteura 5, 02-093 Warsaw, Poland}

\eads{\mailto{jacopo.de.nardis@ens.fr}, \mailto{milosz.panfil@fuw.edu.pl}}

\begin{abstract}
We {review} the recently introduced thermodynamic form factors for pairs of particle-hole excitations {on finite-entropy states} in the Lieb-Liniger model. We focus on the density operator and we show how the form factors can be used for analytic computations of dynamical correlation functions. We derive a new representation for the form factors and we discuss some aspects of their structure. We rigorously show that in the small momentum limit (or equivalently, on hydrodynamic scales) a single particle-hole excitation fully saturates the spectral sum and we also discuss the contribution from two particle-hole pairs. Finally we show that thermodynamic form factors can be also used to study the ground state correlations and to derive the edge exponents.
\end{abstract}

\newpage
\tableofcontents
\newpage
\section{Introduction}

The study of correlation functions of integrable models have a long history. Indeed, it seems natural to expect a knowledge of correlation functions from models which are exactly solvable. The reality however is more complicated and the question of computing correlation functions turns out to be quite involved. Nowadays exact solvability, or quantum integrability usually refers to the Bethe Ansatz methods of finding eigenstates and eigenenergies. Methods of Bethe Ansatz have been successfully applied to very different physical systems, ranging from condensed matter systems~\cite{KorepinBOOK}  through the AdS/CFT correspondence~\cite{2012_AdS/CFT} to supersymetric gauge theories~\cite{2009_SUSY_BA}.

With correlation functions the situation is different. The building blocks of the correlation functions, form factors or matrix elements of local operators, are difficult to compute. They are known only for some models and for some operators. Examples are the Lieb-Liniger gas~\cite{1990_Slavnov_TMP_82,1742-5468-2007-01-P01008,1742-5468-2011-11-P11017,1751-8121-48-45-454002,2015_Pakaliuk_JPA_48} or the XXZ spin chain~\cite{1999_Kitanine_NPB_554,2007_Castro_Alvaredo_JPA_40}. However, even knowing the form factors, to get a correlation function a spectral sum has to be performed which is an even more difficult challenge. The origin of these difficulties is in the same time the reason of interest in integrable models. 

The models under consideration are strongly correlated, which among other things, means that form factors of physical operators, like density or field operators, computed between two arbitrary eigenstates are generally non-zero. Of course it might happen that symmetry forces the form factor to be zero, but unless this is the case, generally they are not zero. This has to be compared with weakly correlated theories, where form factors between two arbitrary eigenstates are generally zero. This way of thinking of correlated systems goes back to the notion of orthogonality catastrophe~\cite{1967_Anderson_PRL_18}. 

Therefore strong correlations in these theories show up in two ways. The form factors are complicated, rather than simple functions of the eigenstates. And in computing the correlation functions, when the spectral sum must be evaluated, one has to sum over a huge number of eigenstates. Some methods were developed to tackle these problems.

In the first approach, instead of studying the full correlation functions, certain special features of them were extracted. For example the ground states of gapless 1D quantum models exhibit critical properties: the space-time correlation functions decay as power laws. A theory of 1 dimensional quantum liquids, the Luttinger theory, predicts a general asymptotic structure of correlation functions~\cite{1981_Haldane_PRL_47,2004_Cazalilla_JPB_37}. By carefully analysing the excitations responsible for that structure, it was possible to derive the asymptotic of the space-time correlation functions directly from the integrable models~\cite{1742-5468-2011-12-P12010,2011_Kozlowski_JSTAT_P03019,1742-5468-2012-09-P09001,1706.09459}. This has an advantage over a universal approach of the Luttinger liquid theory, it allows to go beyond the field theory predictions and also to fix non-universal, model dependent, constants appearing in the expression for the correlation function. These constants can also be fixed without computing the correlation function, but only by computing the effects of the excitations which are responsible for the corresponding part of correlation function~\cite{2011_Shashi_PRB_84,2012_Shashi_PRB_85_1,1742-5468-2012-09-P09001}. 

The second approach, which goes under the name of \texttt{ABACUS}~\cite{2009_Caux_JMP_50}, is to compute correlation functions with the help of a computer. This direct approach relies on knowledge of exact eigenstates, eigenenergies and form factors at finite system size.   The role of the computer is to evaluate the spectral sum term by term. As it is impossible to sum over the whole Hilbert space, the key behind the \texttt{ABACUS} is to organize the sum such that the more important excitations are summed first. This approach turned out to be very successful in providing both qualitative and quantitative predictions for the correlation functions in various physical situations, examples being the ground state~\cite{2006_Caux_PRA_74} and thermal correlation in some range of temperatures~\cite{PhysRevA.89.033605}.

Another way is the so-called quantum transfer matrix approach \cite{0305-4470-37-31-001}. There the correlation functions are given by a series over matrix elements of a time-dependent quantum transfer matrix rather than the Hamiltonian. Recently this approach has brought also new results for the dynamical correlation functions at finite temperature in lattice integrable models \cite{1742-5468-2017-11-113106}.  

Finally methods from integrable field theories were also successfully implemented in the study of dynamical correlations in spin chains \cite{PhysRevB.78.100403,1742-5468-2008-03-P03012,1742-5468-2010-11-P11012}. 

In the work~\cite{Smooth_us} we started a new approach based on the thermodynamic Bethe ansatz. Motivated by a computation of the thermodynamic limit of the ground state form factors~\cite{2012_Shashi_PRB_85_1}  we considered the thermodynamic limit of form factors for a generic (but non-critical) state. By non-critical state we mean a state in which correlation functions do not exhibit critical behavior, {\emph{e.g.} the decay of spatial correlations is exponential}. This excludes for example, the ground state of the Lieb-Liniger model. Focusing on non-critical states allowed us to derive a general expression for the thermodynamic limit of the form factors of the density operators. 

The thermodynamic form factors, however simpler in usage that their finite-size versions are still too complicated for analytical computation of the whole spectral sum. However certain simplifications happen when we consider correlation functions at small momentum. In~\cite{SciPostPhys.1.2.015} we found a simple expression for the low-momentum static structure factor of the Lieb-Liniger model for a generic non-critical state. It turned out that this expression works also for the ground state, where in principle our approach is not valid. Moreover we show there that at low momentum only a single particle-hole excitation is necessary in order to saturate the full spectral sum. This surprising result is related to the so-called generalized hydrodynamics theory \cite{arXiv.1605.09790,Ben}, where the dynamics of any integrable system at large scale is given by the ballistic motion of single particle excitations.    

In this work we review our studies of the thermodynamic form factors of the density operator, focusing on various aspects of them. After the introductory part, the article is divided in few sections which, to certain extent can be read independently. In section \ref{sec:entropyFF} we properly define the thermodynamic form factors. In section~\ref{sec:new_expression} we present and derive a new representation for the thermodynamic limit of the form factors of the density operator. Section~\ref{sec:small_k} focuses on the density-density correlation function in the small momentum limit. We present a new derivation of the detailed balance relation and confirm predictions of the generalized hydrodynamics. In section~\ref{two_ph_and_more} we connect the expansion of the correlation function in the number of particle-hole pairs with an expansion in powers of momentum. In section~\ref{sec:dressing} we show the main difference between the form factors for critical and non-critical states and we describe in which situations we can still use the non-critical form factors to study critical systems. Based on this result, in section~\ref{sec:edge} we derive the edge singularities of the ground state dynamic structure factor in the small momentum limit. In section \ref{sec:ABACUS} we show that correlation functions in the small momentum limit computed with these new form factors agrees with the results obtained from the \texttt{ABACUS} algorithm. In Appendix~\ref{app:averaging_state} we display some open questions regarding the structure of the form factors.  Some more technical or longer computations are placed in Appendix~\ref{app:derivation}. Appendix~\ref{sec:freddet} discusses numerical evaluation of the form factors.

\section{The Lieb-Liniger model}

The Hamiltonian of the Lieb-Liniger model is~\cite{1963_Lieb_PR_130_1, KorepinBOOK}
\ea {
  H = \int_0^L {\rm d} x \left(-\psi^{\dagger}(x)\frac{\partial^2}{\partial x^2}\psi(x) + c \ \psi^{\dagger}(x)\psi^{\dagger}(x)\psi(x)\psi(x) \right),
}
where $L$ is the length of the system, which we assume to be large, and $\psi(x)$ is the canonical Bose field
\ea {
  [\psi(x), \psi^{\dagger}(y)] = \delta(x-y).
}
The Hamiltonian describes a system of bosons on a line and it is a paradigmatic example of a system of interacting bosons on the continuum, also experimentally relevant for cold atomic physics  \cite{1998_Olshanii_PRL_81,2008_Amerongen_PRL_100,Bouchoule,PhysRevA.91.043617,PhysRevLett.115.085301,PhysRevA.96.033624,2017arXiv170800031P}. Notable importance in the past years have had also its non-equilibrium properties \cite{1742-5468-2016-6-064009,Meinert945,PhysRevLett.113.035301}, especially after a quantum quench \cite{1712.04642,2012_Shashi_PRB_85,PhysRevA.89.013609,2014_DeNardis_PRA_89,PhysRevA.91.051602,1751-8121-48-43-43FT01,1367-2630-18-4-045010,
1710.11615,1742-5468-2016-6-064009,1751-8121-50-50-505003,1712.05262}.

We consider a finite but very long system of length $L$ with periodic boundary conditions. Eigenstate are parametrized by a set of $N$ momenta or rapidities $\lambda_j$ which solve the Bethe ansatz equations \eqref{bethe}. We focus on the repulsive regime $c  \geq 0$, where the $\lambda_j$ are all real parameters, in contrast  to the attractive regime $c<0$ where they can form bound states \cite{PhysRevLett.98.150403,PhysRevLett.116.070408,SciPostPhys.1.1.001}.  In the thermodynamic limit $N,L \to \infty$ we can introduce a function $\rho(\lambda)$ which specifies the density of particles with rapidity $\lambda$. The density of particles $\rho(\lambda)$ is related to the filling function $\vartheta(\lambda)$ through the total density density of states $\rho_t(\lambda)$
\ea {
  \vartheta(\lambda) = \frac{\rho(\lambda)}{\rho_t(\lambda)}. \label{rho_particle}
}
The total density function obeys the following integral equation
\ea {
  \rho_t(\lambda) = \frac{1}{2\pi} + \int_{-\infty}^{\infty}{\rm d}\lambda'\, \frac{\vartheta(\lambda')}{2\pi} K(\lambda-\lambda') \rho_t(\lambda'), \label{rho_total}
}
where the kernel $K(\lambda)$ depends on the interaction strength $c$ and is given by
\ea {
  K(\lambda) = \frac{2c}{c^2 + \lambda^2}.
}
For convenience we introduce also a hole density function
\ea {
  \rho_h(\lambda) = \rho_t(\lambda)(1 - \vartheta(\lambda)). \label{rho_hole}
}

Let $|\vartheta\rangle$ denotes a macroscopic eigenstate of the Lieb-Liniger Hamiltonian (whose definition will be more extensively explained in section \ref{sec:entropyFF}) in the thermodynamic limit characterized by a filling function $\vartheta(\lambda)$ taking values in $[0,1]$. 
The particle density, momentum and energy are the consecutive moments of the density~$\rho(\lambda)$
\ea {
  n \equiv  \frac{N}{L} &= \int_{-\infty}^{\infty}{\rm d}\lambda\, \rho(\lambda),\\
  \frac{P}{L} &= \int_{-\infty}^{\infty}{\rm d}\lambda\, \lambda \rho(\lambda),\\
  \frac{E}{L} &= \int_{-\infty}^{\infty}{\rm d}\lambda\, \lambda^2 \rho(\lambda).
}
In the following we set the total density to unit value $n=1$.  We give few examples of the filling function for physically interesting states. The filling function of the ground state corresponds to a Fermi~sea
\ea {
  \vartheta(\lambda) = \begin{cases}
    1, \quad -q \geq \lambda \geq q,\\
    0, \quad \textrm{otherwise},
  \end{cases}
}
where $q$ plays a role of the Fermi momentum. The ground state is an archetypical, and the most important, critical state. Other interesting example of a critical state is the split Fermi sea state introduced in~\cite{2014PhRvA..89c3637F}. The filling function for the finite temperature state \cite{1969_Yang_JMP_10} is
\ea { \label{varthetathermal}
  \vartheta(\lambda) = \frac{1}{1 + \exp(\beta (\varepsilon(\lambda) - \mu))},
}
where $\varepsilon(\lambda)$ is the dressed energy \eqref{drenergy} and $\mu$ the chemical potential. Other distributions of interest are the generalized Gibbs ensemble (GGE) states \cite{PhysRevLett.106.140405,Pasquale-ed,1742-5468-2016-6-064007,1742-5468-2016-6-064002}, which are nothing else than generalization of the thermal state \eqref{varthetathermal}, with $\beta \to \beta(\lambda)$ with $\beta(\lambda)$ a positive function~\cite{PhysRevB.84.212404,SciPostPhys.3.3.023}
\ea { \label{varthetaGGEl}
  \vartheta(\lambda) = \frac{1}{1 + \exp(\beta(\lambda) (\varepsilon(\lambda) - \mu))}.
}
Contrary to the ground state, the filling function for finite temperature and GGE states is a smooth function of $\lambda$ for any $\beta(\lambda)<\infty$. We remark that since the total density of the gas have to be finite, the functions $\varepsilon(\lambda)$ and $\beta(\lambda)$ must be such that $\int d\lambda \vartheta(\lambda)  < \infty$. 

The excited states around $|\vartheta\rangle$ are created by making a number of particle-hole pairs in the filling function. An $m$ particle-hole excited state we denote $|\vartheta; \mathbf{p},\mathbf{h}\rangle$, where $\mathbf{p} = \{ p_j\}_{j=1}^m$ and $\mathbf{h} = \{ h_j \}_{j=1}^m$.  Sets $\mathbf{p}$ and $\mathbf{h}$  specify the particle-hole content of the excited state. The particle density for such an excited state is
\ea {
  \rho(\lambda; \mathbf{p}, \mathbf{h}) = \rho(\lambda) + \frac{1}{L}\sum_{j=1}^m \left(\delta(\lambda - p_j) - \delta(\lambda - h_j) \right)  - \frac{1}{L} \partial_\lambda\rho(\lambda)\left (  \frac{ F(\lambda|  \mathbf{p}, \mathbf{h})}{\rho_t(\lambda)}\right) +\mathcal{O}(L^{-2}),
}
where the backflow function $F(\lambda|  \mathbf{p}, \mathbf{h}) = \sum_{j=1}^m \left(  F(\lambda|  p_j) -   F(\lambda|  h_j) \right)$ is defined below. 
There are also more general excited states with different number of particles and holes but the form factors of the density operators vanishes for such states. An excited state has relative ({with respect to $|\vartheta\rangle$}) energy and momentum given by
\ea {
  k(\vartheta; \mathbf{p}, \mathbf{h}) &= \sum_{j=1}^m k(p_j) -  k(h_j),\label{k}\\
  \varepsilon(\vartheta; \mathbf{p}, \mathbf{h}) &= \sum_{j=1}^m \varepsilon(p_j) - \varepsilon(h_j).\label{omega}
}
 {Functions $k(\lambda)$ and $\varepsilon(\lambda)$ are the dressed momentum and energy and are given by} 
\ea {
k(\lambda) &= \lambda + \int_{-\infty}^{\infty} {\rm d}\alpha F(\alpha|\lambda) \vartheta(\alpha), \\
\varepsilon(\lambda) &= \lambda^2 + \int_{-\infty}^{\infty} {\rm d}\alpha F(\alpha|\lambda) \vartheta(\alpha)  (2 \alpha).\label{drenergy}
}
Here $F(\alpha| \lambda )$ is the backflow or shift function obeying
\ea {\label{backflow}
 2\pi F(\lambda|\alpha) \!=\! \theta(  \lambda - \alpha) +\! \int_{-\infty}^{\infty}\!\!{\rm d}\lambda' \vartheta(\lambda')F(\lambda'|\alpha)K(\lambda - \lambda'),
}
where
\ea {
  \theta(\lambda) = 2\arctan\left(\frac{\lambda}{c}\right).
}
Notice that the total density of states $\rho_t(\lambda)$ is the derivative of the dressed momentum 
\begin{equation}
k'(\lambda) = 2 \pi \rho_t(\lambda).
\end{equation}

In this work we are concerned with the density-density correlation functions, also known as a dynamic structure factor, DSF, in the thermodynamic limit at fixed total density $n$. The density operator is
\ea {
  \hat{\rho}(x) = \psi^{\dagger}(x)\psi(x),
}
and its time evolution in the Heisenberg picture is given by the Lieb-Liniger Hamiltonian
\ea {
  \hat{\rho}(x,t) = e^{iHt}\hat{\rho}(x) e^{-iHt}.
}
The dynamical density-density correlation function of the state $|\vartheta\rangle$ in the thermodynamic limit is defined as
\ea {
  S_{\hat{\rho}}(x,t) = \langle \vartheta|\hat{\rho}(x,t) \hat{\rho}(0,0)|\vartheta\rangle.
}
Its Fourier transform, the dynamic structure factor, is given by
\ea {
  S_{\hat{\rho}}(k,\omega) =  \int_{-\infty}^{\infty} {\rm d}x \int_{-\infty}^{\infty} {\rm d}t\, e^{i(kx - \omega t)} S(x,t).
}
In the spectral representation this can be written as a sum over a generic number of pairs of particle-hole excitations on the reference state $|\vartheta\rangle$~\cite{Smooth_us}
\ea { \label{S_expansion}
  S_{\hat{\rho}}(k,\omega) = \sum_{m \geq 1 } S^{m\text{ph}}_{\hat{\rho}}(k,\omega)   ,
}
where the contribution from $m$ particle-hole pairs is given by
\ea { \label{mph_contribution}
 S^{\text{mph}}_{\hat{\rho}}(k,\omega) =     \frac{(2\pi)^2}{(m!)^2}  \fint_{-\infty}^{\infty} {\rm d}{\mathbf p}_{m} {\rm d}{\mathbf h}_{m} |\langle \vartheta| \hat{\rho}(0)|\vartheta, \mathbf{h} \to  \mathbf{p}  \rangle|^2 \delta(k - k(\mathbf{p}, \mathbf{h}))\delta(\omega - \omega(\mathbf{p}, \mathbf{h})).
}
Here the integration measure is defined~as
\ea {
  {\rm d}\mathbf{p}_m{\rm d}\mathbf{h}_m = \prod_{j=1}^m {\rm d}p_j {\rm d}h_j\, \rho(h_j) \rho_h(p_j),
}
and the finite part integral is defined~as
\ea { \label{finite_part_integral}
  \fint_{-\infty}^{\infty} {\rm d}h  f(h) = \lim_{\epsilon\rightarrow 0^+}\int_{-\infty + i \epsilon}^{\infty + i \epsilon} {\rm d}h\,  f(h) + \pi i \underset{h=p}{\rm res}   f(h).
}
The finite part integral appears because the thermodynamic form factors $ |\langle \vartheta| \hat{\rho}(0)|\vartheta, \mathbf{h} \to  \mathbf{p}  \rangle|$ have a single pole when $p_j$ coincides with $h_k$, known as \textit{kinematic poles}. In the next section we review how to properly define the thermodynamic form factors on a generic non-critical thermodynamic eigenstate $| \vartheta\rangle$.

\section{Entropy of states and thermodynamic form factors}\label{sec:entropyFF}
In this section we review the definition of the thermodynamic form factors and formula for the form factor of the density operator orginally presented in~\cite{Smooth_us}. In order to define the thermodynamic form factors $\langle \vartheta|\hat{\rho}(0)|\vartheta; \mathbf{h}\rightarrow \mathbf{p}\rangle$ we proceed in the following way. We consider a finite system with periodic boundary conditions. Then the eigenstates of the Hamiltonian are parametrized by a set of quantum numbers $\{I_j\}_{j=1}^N$, where $N$ is the number of particles. Physical quantities are expressed in terms of the rapidities which are related to quantum numbers through Bethe equations
\ea { \label{bethe}
  \lambda_j = \frac{2\pi}{L} I_j + \sum_{k \neq j = 1}^N \theta(\lambda_j - \lambda_k), \quad \quad j=1,\ldots, N.
}
In the thermodynamic limit $N,L\rightarrow \infty$ with $N/L=n$ fixed, the rapidities are described through their density function, the particle density $\rho(\lambda)$ which can be formally defined~as
\ea {
  \rho(\lambda) = \lim_{N,L \to \infty}  \frac{1}{L}\sum_{j=1}^N \delta(\lambda - \lambda_j).
}
The particle density can be used to compute the filling function $\vartheta(\lambda)$ which then specifies a thermodynamic state $|\vartheta\rangle$.  Let $\{I_j\}_{j=1}^N$ be a set of quantum numbers specifying a Bethe state such that in the thermodynamic limit its filling function is given by $\vartheta(\lambda)$. There are many choices of quantum numbers leading to the same filling function, thus to the same thermodynamic state $|\vartheta\rangle$. Their number is $\exp S[\vartheta]$ where $S[\vartheta]$ is the extensive Yang-Yang entropy~\cite{1969_Yang_JMP_10}
\begin{equation}\label{yangyang}
S[\vartheta] = L \int_{-\infty}^{\infty} {\rm d}\lambda \left( \rho_t(\lambda) \log \rho_t(\lambda) -    \rho_h(\lambda) \log \rho_h(\lambda)  -  \rho(\lambda) \log \rho(\lambda) \right).
\end{equation}
We define the normalized thermodynamic state as
\ea { \label{thermstate}
  |\vartheta\rangle = \lim_{N,L \to \infty} \exp\left(- \frac{1}{2} S[\vartheta] \right)\sum_{\{I_j\}} |\{I_j\}\rangle,
}
where the summation is over all the $e^{S[\vartheta]}$ microscopic states with the same $\vartheta(\lambda)$ in the thermodynamic limit. Notice that this definition was introduced also in the context of the Quench Action approach \cite{QA_Caux}. 

The density operator $\hat{\rho}(x)$, like all other local operators, acts almost diagonally, which means that for a given excitation and a choice of $\{I_j\}$ the set $\{J_j\}$ is basically fixed up to pairs of particle-hole excitations $\text{ph}$ which correspond to changing the values of some quantum numbers: $ \text{ph} \equiv \{ I_i \to J_i \}_{i=1}^m $, for a number $m$ small compared to the system size $L$. This implies that the only non-zero matrix elements are the one where right and left states are characterized by the same filling function in the thermodynamic limit . The thermodynamic form factors then reads
\ea { \label{eqthesame}
  \langle \vartheta|\hat{\rho}(0)|\vartheta,  \mathbf{h} \to \mathbf{p} \rangle &= \lim_{L,N\to\infty} \exp\left(-\frac{1}{2} (S[\vartheta]+S[\vartheta,\text{ph}])\right)L^{m}\sum_{\{ I_j\}} \langle\{ I_j\}|\hat{\rho}(0)|\{ I_j+\text{ph}\} \rangle,
}
where $S[\vartheta,\text{ph}] = S[\vartheta] + \mathcal{O}(L^0)$ is the entropy of the right state. In order to take the thermodynamic limit we choose a $\text{ph}$ such that in the thermodynamic limit there is a set of finite particle-hole pairs $ \mathbf{h} \to \mathbf{p} $ in the space of rapidities $\lambda$. For each hole $h$ and particle $p$ there are $\sim L$ possibilities for $\text{ph}$, which is the reason why we multiply times $L^{m}$ in \eqref{eqthesame}. This degeneracy will be taken into account also by multiplying the square of the thermodynamic form factors times the densities of state $\rho_h(p)\rho(h)$. Notice that this picture changes drastically when the quantum numbers $\{ I_j^0\}$ describe the ground state or a state with a finite discontinuity in $\vartheta(\lambda)$ in the thermodynamic limit. In this case the finite size form factors decay with a non-integer power of $L$ \cite{2012_Shashi_PRB_85_1,1742-5468-2012-09-P09001}.

In our computations we assumed that each element of the sum is essentially the same (for large system size) and therefore
 \ea {
 \sum_{\{ I_j\}} \langle\{ I_j\}|\hat{\rho}(0)|\{ I_j+\text{ph}\} \rangle = \exp\left(S[\vartheta]\right) \langle\{ I_j^0\}|\hat{\rho}(0)|\{ I_j^0+\text{ph}\},
}
where $\{ I_j^0\}$ is any of the states. Finally using \eqref{thermstate} we obtain the definition of the thermodynamic form factors
\ea {
  \langle \vartheta|\hat{\rho}(0)|\vartheta,  \mathbf{h} \to \mathbf{p} \rangle &= \exp\left(\frac{1}{2}\delta S[\vartheta, \phset]\right) \lim_{L,N\to\infty} L^{m}\langle\{ I_j^0\}|\hat{\rho}(0)|\{ I_j^0+\text{ph}\}\rangle.
} 
The differential entropy is defined as, see eq. \eqref{diff_entropy}, 
\ea {
  \delta S[\vartheta, \phset] = S[\vartheta,\phset] - S[\vartheta].
}
The state $|\{I_j^0\}\rangle$ is called the averaging state and can be any state with $\vartheta(\lambda)$ in the thermodynamic limit.  Computations of the thermodynamic limit of the form factors simplify after a convenient choice of the averaging state. In our work \cite{Smooth_us} we have chosen a uniform averaging state, meaning that for each interval $[\lambda, \lambda + {\rm d}\lambda]$ there are $\rho(\lambda){\rm d}\lambda$ uniformly distributed rapidities. The role of the averaging state and consequences of different choices on the computation of thermodynamic limit of the form factors are discussed in the appendix~\ref{app:averaging_state}.

\subsection{Thermodynamic form factors} \label{sec:TL}

We recall here the formula for the form factors of the density operator $\hat{\rho}(x)$ acting in position $x=0$ in the thermodynamic limit. This formula was derived in~\cite{Smooth_us} and after a small change of notation that we describe later reads
\ea { \label{FF1}
  |\langle \vartheta| \hat{\rho}(0)&|\vartheta, \mathbf{h} \to \mathbf{p} \rangle | = \mathcal{A}(\vartheta, \phset) \mathcal{D}(\vartheta, \phset) \exp\left(\mathcal{B}(\vartheta, \phset)\right) ,
}
where\footnote{In the formula for the form factors presented in~\cite{Smooth_us} and~\cite{SciPostPhys.1.2.015} there was a misprint and factor $1/2$ in front of the differential entropy $\delta S[\vartheta;\phset]$ was missing.  Morevoer in~\cite{SciPostPhys.1.2.015} the last factor of $\mathcal{B}(\vartheta, \{ p_j, h_j\})$ was also missing. However, since both contributions scale like $k$, where $k$ is the momentum of the excited state, these factor do not contribute in the small momentum limit and do not change the results of ~\cite{SciPostPhys.1.2.015}.}
\ea {
 \mathcal{A}(\vartheta, \phset) =& \prod_{k=1}^m \left[\frac{F(h_k)}{\left(\rho_t(p_k)\rho_t(h_k) \right)^{1/2}} \frac{\pi \tilde{F}(p_k)}{\sin \pi\tilde{F}(p_k)} \frac{\sin \pi\tilde{F}(h_k)}{\pi \tilde{F}(h_k)} \right] \nn\\
 &\times  \prod_{i,j=1}^m \left[\frac{(p_i - h_j + ic)^2}{(h_{i,j} + ic)(p_{i,j}+ic)} \right]^{1/2}\frac{\prod_{i<j=1}^m h_{i,j} p_{i,j}}{\prod_{i,j=1}^m (p_i-h_j)},\label{FF_A} \\
 \mathcal{B}(\vartheta, \phset) =&  -\frac{1}{4}\int_{-\infty}^{+\infty} {\rm d}\lambda {\rm d}\lambda' \left(\frac{\tilde{F}(\lambda) - \tilde{F}(\lambda')}{\lambda -\lambda'} \right)^2 - \frac{1}{2} \int_{-\infty}^{+\infty} {\rm d}\lambda {\rm d}\lambda' \left(\frac{\tilde{F}(\lambda)\tilde{F}(\lambda')}{(\lambda-\lambda'+ic)^2} \right)\nn\\
 &+ \sum_{k=1}^m \pint_{-\infty}^{+\infty} {\rm d}\lambda \frac{\tilde{F}(\lambda)(h_k-p_k)}{(\lambda-h_k)(\lambda-p_k)} + \int_{-\infty}^{+\infty} {\rm d}\lambda \frac{\tilde{F}(\lambda)(p_k-h_k)}{(\lambda-h_k+ic)(\lambda-p_k+ic)} \nn\\
 &+ \frac{1}{2} \delta S[\vartheta;\phset] + \frac{1}{2}\int {\rm d}\lambda\, \vartheta(\lambda)  {F}'(\lambda) \pi   {F}(\lambda) \cot(\pi  {F}(\lambda)),\label{FF_B} \\
 \mathcal{D}(\vartheta, \phset) =& \,\frac{c}{2 } \,{\rm det}_{i,j=1}^m\left( \delta_{ij} + W_{ij}\right) \frac{{\rm Det}(\boldsymbol{1} - \hat{A})}{{\rm Det}(\boldsymbol{1}-\hat{K}_{\vartheta})}.\label{FF_D}
}
We now explain the main ingredients of this formula. The  function $\vartheta(\lambda)$ is the filling function characterizing a state and $\{p,h\}_{j=1}^m$ describe  particle-hole excitations over this state. $F(\lambda) = F(\lambda| \mathbf{p}, \mathbf{h})$  is the back-flow function~\eqref{backflow}. To shorten the formula we also use a rescaled back-flow $\tilde{F}(\lambda) = \vartheta(\lambda)F(\lambda)$.
Function $\delta S[\vartheta;\phset]$ is a differential entropy defined as
\ea {
  \delta S[\vartheta;\phset] = \int_{-\infty}^{\infty}{\rm d}\lambda\, s[\vartheta;\lambda] \frac{\partial}{\partial \lambda}\left(\frac{F(\lambda|\phset)}{\rho_t(\lambda)} \right), \label{diff_entropy}
}
where $s[\vartheta;\lambda]$ is the entropy density~\cite{1969_Yang_JMP_10} expressible through the density functions (see eq.~\eqref{yangyang})
\ea {
  s[\vartheta;\lambda] = \rho_t(\lambda)\log \rho_t(\lambda) - \rho(\lambda)\log\rho_p(\lambda) - \rho_h(\lambda)\log\rho_h(\lambda).
}

The first determinant of $\mathcal{D}(\vartheta, \phset)$ is of a square matrix with the same size as the number of particle-hole pairs. Its matrix elements $W(h_i,h_j)$ are defined as solutions to the following linear integral equation
\ea{ \label{eq_W}
  W(h_i, \lambda) - \pint_{-\infty}^\infty {\rm d}\alpha\, W(h_i,\alpha) \tilde{a}(\alpha) \left(K(\alpha - \lambda) - \frac{2}{c} \right) = b_i\left(K(h_i - \lambda) - \frac{2}{c} \right)   ,
}
with the vector $b_i$ given by
\ea { \label{b_i}
  b_i = - \frac{\resatilde^{[\mathbf{p}, \mathbf{h}]}}{\vartheta(h_i) F(h_i)},
}
where $\tilde{a}^{[\mathbf{p}, \mathbf{h}]}(\lambda)$ is a function defined below and  $\resatilde^{[\mathbf{p}, \mathbf{h}]}$ is its residue at $\lambda = h_i$
\ea { \label{a_i}
  \resatilde^{[\mathbf{p}, \mathbf{h}]} = \lim_{\lambda\rightarrow h_i} (\lambda - h_i)\tilde{a}^{[\mathbf{p}, \mathbf{h}]}(\lambda).
}
The two other determinants of $\mathcal{D}(\vartheta, \phset)$ are Fredholm determinants. The kernel $\hat{A}$ is
\ea {\label{kernel_A}
  A(\lambda,\lambda') = \tilde{a}^{[\mathbf{p}, \mathbf{h}]}(\lambda)\left(K(\lambda-\lambda') - \frac{2}{c}\right),
}
where
\ea { 
  \tilde{a}^{[\mathbf{p}, \mathbf{h}]}(\lambda)  =  \frac{\sin[ \pi \vartheta(\lambda) F(\lambda)]}{2\pi \sin [ \pi F(\lambda)]} \prod_{k=1}^m  \nn & \left(  \frac{  p_k - \lambda}{ h_k - \lambda}
 \sqrt{\frac{K(p_k -\lambda)}{K(h_k - \lambda)} }\right)
e^{ -\frac{c}{2} \pint_{-\infty}^{\infty} {\rm d}
    \lambda'  \frac{\vartheta(\lambda')F(\lambda') K(\lambda' - \lambda)}{\lambda' - \lambda} }.\label{atilde}
}
The  {}{operator} $\boldsymbol{1}$ represents the identity $\boldsymbol{1}(\lambda,\lambda') = \delta(\lambda-\lambda')$ and 
\begin{equation}
\hat{K}_{\vartheta} (\lambda, \lambda')= K(\lambda- \lambda') \frac{\vartheta(\lambda')}{2 \pi}.
\end{equation}
Comparing with expressions presented in~\cite{Smooth_us,SciPostPhys.1.2.015} in the present formula we changed the sign of $\tilde{a}^{[\mathbf{p}, \mathbf{h}]}(\lambda)$ which leads to: a minus sign in the Fredholm determinant ${\rm Det}(\boldsymbol{1} - \hat{A})$, a changes of sign in the integral equation~\eqref{eq_W} specifiying matrix elements $W(h_i,h_j)$, and a change of sign in expression~\eqref{b_i} for $b_i$. This last change is cancelled, because computation of the residue~\eqref{a_i} gives an extra minus sign.

In the expression for the form factors principal value integrals appear. We define them as
\ea { \label{Pint}
  \pint_{a}^{b} {\rm d}\lambda \frac{f(\lambda)}{\lambda - c} = \lim_{\epsilon\rightarrow 0} \left( \int_{a}^{c-\epsilon} {\rm d}\lambda \frac{f(\lambda)}{\lambda - c} + \int_{c+\epsilon}^{b} {\rm d}\lambda \frac{f(\lambda)}{\lambda - c}\right), \qquad a < c < b.
}
In this notation
\ea {
  \pint_{-a}^a \frac{{\rm d}\lambda}{\lambda} = 0.
}
The principal value integrals appear because of the need to regularize the thermodynamic limit of certain terms in the form factors. In appendix~\ref{app:averaging_state} we recall how the form of the principal value integral is connected with the choice of the averaging state. 

\begin{figure}
  \center
  \includegraphics[width=0.85\textwidth]{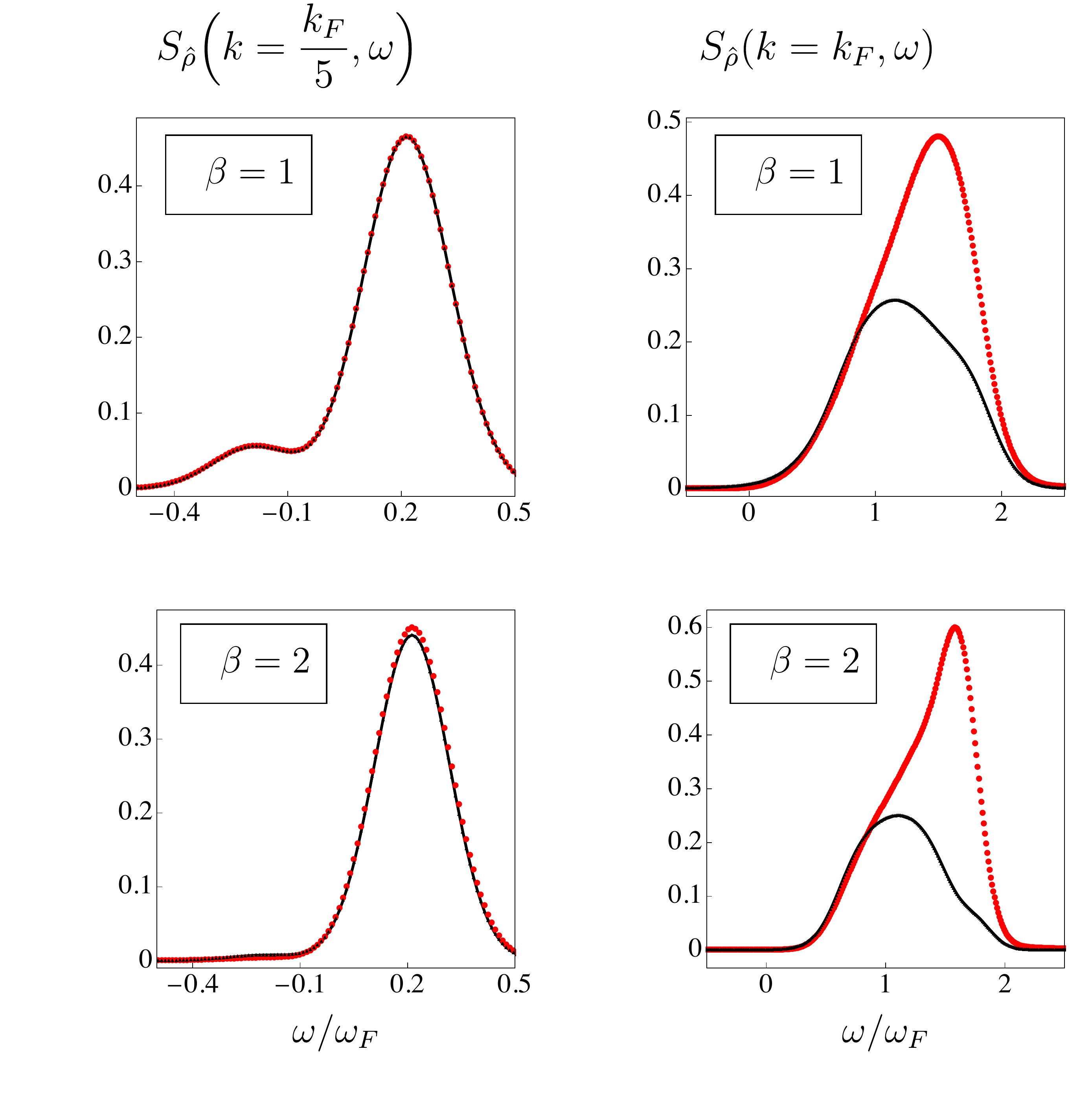}
  \caption{Dynamical structure factor (DSF) $S_{\hat{\rho}}(k,\omega)$ smoothed in energy $\omega$ by convolving it with a Gaussian distribution of unit variance and evaluated at $c=4$, on thermal states with two different inverse temperatures $\beta=1$ and $\beta=2$ and unitary density $n=1$. Red data are obtained numerically by the \texttt{ABACUS} algorithm while black continuous data are obtained by evaluations of the single particle-hole DSF $S^{\text{1ph}}_{\hat{\rho}}(k,\omega)$ \eqref{single_particle_DSF}. At $k/k_F=0.2$ (with $k_F = \pi n$ and $\omega_F = k_F^2$) the single particle-hole contribution completely saturates the full dynamical sum and $S_{\hat{\rho}}(k,\omega) \simeq S^{\text{1ph}}_{\hat{\rho}}(k,\omega)$, while at large $k= k_F$ the two and higher particle-hole contributions are necessary in order to compute the full DSF. For more details see section \ref{sec:ABACUS}.  }
  \label{fig:DSF}
\end{figure}

\section{A new expression for the thermodynamic form factors}\label{sec:new_expression}

We will show now that the last part of the form factors can be written in a simpler form
\ea { \label{eq:newD}
  \mathcal{D}(\vartheta, \phset) =& \,\frac{ {\rm det}_{i,j=1}^m\left( A_{ij} + B_{ij}\right)  }{{\rm Det}(\boldsymbol{1} + L^{[\mathbf{p}, \mathbf{h}]})  {\rm Det}\Big(\boldsymbol{1}- \hat{K}_{\vartheta}\Big)}.}
which is given only in terms of the ``generalized particle-hole resolvent'' $L^{[\mathbf{p}, \mathbf{h}]}$, see eq. \eqref{Lph_int_eq}. This new expression is also simpler from the numerical evaluation point of view, as only one extra solution of integral equation is needed  in order to determine the full form factor. Moreover we will see later that it leads to a much simpler expression of the single particle-hole form factor \eqref{singlep_h_FF} and of its small momentum limit.   {In this section we will present the main idea behind the derivation of this new expression. Few more technical steps are relegated to the appendix~\ref{app:derivation}}.
The matrix elements $A_{ij}$, $B_{ij}$, such that $A_{ij} + B_{ij} = \delta_{ij} +  W(h_i,h_j)$, can be written in terms of ``generalized particle-hole thermodynamic functions''
\ea {\label{newW}
  A_{ij} &=  \delta_{ij}  - \frac{\resatilde^{[\mathbf{p}, \mathbf{h}]}}{\vartheta(h_i)F(h_i)}\left[ \lim_{\lambda \to h_j} \frac{L^{[\mathbf{p}, \mathbf{h}]}(h_i,\lambda)}{\tilde{a}^{[\mathbf{p}, \mathbf{h}]}(\lambda)} \right],\\
  B_{ij} &= \frac{\resatilde^{[\mathbf{p}, \mathbf{h}]}}{\vartheta(h_i) F(h_i)} 2\pi \rho_t^{[\mathbf{p}, \mathbf{h}]}(h_i) 2\pi \rho_t^{[\mathbf{p}, \mathbf{h}]}(h_j).
}
which we will soon define.  Notice that we have 
\begin{equation}\label{detAzero}
 {\rm det}_{i,j=1}^m \left(A_{ij} \right) = 0,
\end{equation}
as we expect from the finite size expression of the form factors, see Appendix~\ref{sec:constant_alpha}.
 In the course of the computation we will refer to the small momentum limit of the form factors which we studied in~\cite{SciPostPhys.1.2.015}. By small momentum limit we mean computing form factors between two states $|\vartheta\rangle$ and $|\vartheta;  h \to  p\rangle$ such that the momentum $k$~\eqref{k} between the two states is small. In section~\ref{sec:small_k} we give more details on the small momentum limit. For now we recall that in that case it was useful to introduce the resolvent of the kernel~$\hat{K}_{\vartheta}$
\ea {
  (\boldsymbol{1} + \hat{L}_{\vartheta})\left(\boldsymbol{1} - \hat{K}_{\vartheta} \right) = \boldsymbol{1}  = \left(\boldsymbol{1} - \hat{K}_{\vartheta} \right)(\boldsymbol{1} + \hat{L}_{\vartheta}),
}
which obeys the following integral equation
\ea { \label{L_int_eq}
  \hat{L}_{\vartheta}(\lambda, \lambda') =  \frac{\vartheta(\lambda')}{2\pi} \left(  K (\lambda - \lambda')  + \int_{-\infty}^{\infty} {\rm d}\alpha \hat{L}_{\vartheta}(\lambda, \alpha)  K(\alpha- \lambda')  \right).
}
The resolvent is proportional to the derivative of the shift function \eqref{backflow}, namely
\begin{equation}\label{L-F}
\hat{L}_{\vartheta}(\mu,\lambda) = -\vartheta(\lambda) \partial_\mu F(\lambda|\mu).
\end{equation}
The {introduction of the resolvent $\hat{L}_{\vartheta}$} allows for a number of simplifications which lead to a simple expression for a form factors in the small momentum limit.  
Here we generalize the approach of~\cite{SciPostPhys.1.2.015} to any momentum $k$, using a generalization of the resolvent $\hat{L}_{\vartheta}(\lambda,\lambda')$. The new generalized particle-hole thermodynamic functions reduce to the standard ones in the small momentum limit, i.e. when the position of each particle excitation coincides with the one of its hole.

\subsection{Generalized particle-hole thermodynamic functions}
We define a new kernel
\ea {
 K^{[\mathbf{p}, \mathbf{h}]}(\lambda,\lambda') = K(\lambda,\lambda') \tilde{a}^{[\mathbf{p}, \mathbf{h}]}(\lambda').
}
The resolvent of the new kernel is defined through
\ea { \label{L_ph_def}
  \left(\boldsymbol{1}+L^{[\mathbf{p}, \mathbf{h}]}\right)\left(\boldsymbol{1} -K^{[\mathbf{p}, \mathbf{h}]}\right) = \boldsymbol{1} = \left(\boldsymbol{1} -K^{[\mathbf{p}, \mathbf{h}]}\right)\left(\boldsymbol{1} + L^{[\mathbf{p}, \mathbf{h}]}\right).
}
We can write an integral equation for this generalized resolvent
\ea {
  L^{[\mathbf{p}, \mathbf{h}]}(\lambda, \lambda') =  \tilde{a}^{[\mathbf{p}, \mathbf{h}]}(\lambda') \left(  K (\lambda - \lambda') + \pint_{-\infty}^{\infty} {\rm d}\alpha L^{[\mathbf{p}, \mathbf{h}]}(\lambda, \alpha)  K (\alpha, \lambda') \right). \label{Lph_int_eq}
}
{Function $\tilde{a}^{[\mathbf{p}, \mathbf{h}]}(\lambda)$ has simple poles whenever $\lambda$ coincides with a position of a hole. } Iterating the equation we see that the poles of $\tilde{a}^{[\mathbf{p}, \mathbf{h}]}(\lambda')$ give rise to the poles of $L^{[\mathbf{p}, \mathbf{h}]}(\lambda, \lambda')$. Consequently the particle-hole resolvent has simple poles for $\lambda'$ equal to positions of holes.  We can write an integral equation for the ratio $L^{[\mathbf{p}, \mathbf{h}]}(\lambda, \lambda')/\tilde{a}^{[\mathbf{p}, \mathbf{h}]}(\lambda')$
\ea {
  \frac{L^{[\mathbf{p}, \mathbf{h}]}(\lambda, \lambda')}{\tilde{a}^{[\mathbf{p}, \mathbf{h}]}(\lambda')} = K(\lambda - \lambda') + \pint_{-\infty}^{\infty} {\rm d}\alpha\, \tilde{a}^{[\mathbf{p}, \mathbf{h}]}(\alpha) \frac{L^{[\mathbf{p}, \mathbf{h}]}(\lambda, \alpha)}{\tilde{a}^{[\mathbf{p}, \mathbf{h}]}(\alpha)} K(\alpha, \lambda'). 
}
This form is convenient to study the small momentum limit of the single particle-hole contribution. Under the principal value integral the poles of $\tilde{a}^{[\mathbf{p}, \mathbf{h}]}(\alpha)$ do not matter and we can safely take the small momentum limit of $\tilde{a}^{[\mathbf{p}, \mathbf{h}]}(\alpha)$ which is $\vartheta(\alpha)/(2\pi)$. In turn the integral equation becomes an integral equation for the ratio $2\pi \hat{L}_{\vartheta}(\lambda, \lambda')/\vartheta(\lambda')$, c.f~\eqref{L_int_eq}. Therefore
\ea {
  \frac{L^{[ {p}, {h}]}(\lambda, \lambda')}{\tilde{a}^{[ {p},  {h}]}(\lambda')} \smallk \frac{2\pi \hat{L}_{\vartheta}(\lambda, \lambda')}{\vartheta(\lambda')}.
}
We also define a generalization\footnote{We mean here that $\rho_t^{[\mathbf{p}, \mathbf{h}]}(\lambda)$ obeys a similar integral equation to the one obeyed by $\rho_t(\lambda)$ and in the small momentum limit is equal to $\rho_t(\lambda)$. However we do not attempt here to give a physical interpretation of the relation beween $\rho_t^{[\mathbf{p}, \mathbf{h}]}(\lambda)$ and the total density of states $\rho_t(\lambda)$.\label{footnote}} of the total density of rapidities  $\rho_t^{[\mathbf{p}, \mathbf{h}]}(\lambda)$ which is given in terms of the resolvent $L^{[\mathbf{p}, \mathbf{h}]}$ as
\ea { \label{rho_total_ph}
2\pi \rho_t^{[\mathbf{p}, \mathbf{h}]}(\lambda)  =  1 +  \pint_{-\infty}^{\infty}{\rm d}\alpha\, L^{[\mathbf{p}, \mathbf{h}]}(\lambda,\alpha)  .
}
All these expression depends on the excitations through the $\tilde{a}^{[\mathbf{p}, \mathbf{h}]}(\lambda)$ function. For a single particle-hole pair in the small momentum limit $p \to h$ the generalized functions reduce to the standard ones
\ea {
  \tilde{a}^{[ {p},  {h}]} &\smallk \frac{\vartheta}{2\pi}, \\
  {\tilde{a}_{\text{res}(h)}} ^{[ {p},  {h}]}&\smallk (p - h)\frac{\vartheta(h)}{2\pi},  \\
   K^{[ {p}, {h}]} &\smallk \hat{K}_{\vartheta} ,\\ 
  L^{[ {p}, {h}]} &\smallk {\hat{L}_{\vartheta}},\\
\rho_t^{[ {p}, {h}]}  &\smallk \rho_t.
}

We follow now the same strategy as in the small momentum limit and rewrite the Fredholm determinant ${\rm Det}(\boldsymbol{1}-\hat{A})$ and solve for function $W(\lambda,\lambda')$. The computations are presented in appendix~\ref{app:derivation}. For the Fredholm determinant we find
\ea {
  {\rm Det}\left( \boldsymbol{1} - \tilde{a}^{[\mathbf{p}, \mathbf{h}]}\left(K- \frac{2}{c}\right)\right) = \left(  {1}  + \frac{2 n^{[\mathbf{p}, \mathbf{h}]}  }{c} \right) {\rm Det}\left(  \boldsymbol{1}  -K^{[\mathbf{p}, \mathbf{h}]}\right), 
}
where $n^{[\mathbf{p}, \mathbf{h}]}$ is a generalization of the density of the gas\footnote{The remark of footnote~\ref{footnote} applies also here.}
\ea { \label{n_ph}
 n^{[\mathbf{p}, \mathbf{h}]} =\pint_{-\infty}^{\infty} {\rm d}\lambda \rho_t^{[\mathbf{p}, \mathbf{h}]}(\lambda)  (  2\pi  \tilde{a}^{[\mathbf{p}, \mathbf{h}]}(\lambda)), \quad   n^{[ {p},  {h}]} \smallk n.
}
For the matrix elements $A_{ij}$, $B_{ij}$ we find the result of equation \eqref{newW}, which, in the small momentum limit and for single particle-hole excitation, gives
\ea { 
  A_{11} &=0\\
  B_{11} &\smallk   \frac{2\pi \rho_t(h) \rho(h)}{\hat{L}_{\vartheta}(h,h)},
}
in agreement with the results of~\cite{SciPostPhys.1.2.015}.

\section{Small momentum limit and single particle-hole contribution}\label{sec:small_k}
We review here the results found in \cite{SciPostPhys.1.2.015} for the small momentum limit of the dynamical correlation functions and the single particle-hole contribution. Despite this is a well studied limit, see for example \cite{PhysRevLett.113.015301,PhysRevB.89.100504,1509.08332} very few results are available in this limit for thermal and non-thermal correlations.  
In~\cite{SciPostPhys.1.2.015} we considered a single particle-hole excitation in the limit of small total momentum $k = k(p) - k(h)$ and energy $\omega=  \varepsilon(p) - \varepsilon(h)$. 
By small momentum we mean that $k$ is small compared with another scale set by the interaction parameter $c$. Therefore for large values of $c$ the small momentum limit is actually a valid approximation over a large range of momenta.
The shift function is related to the resolvent $\hat{L}_{\vartheta}$ by 
\begin{equation}\label{backflow_resolvent}
 F(\lambda|p,h) = - \frac{1}{\vartheta(\lambda)} \int_h^p {\rm d}\alpha\,\hat{L}_{\vartheta}(\alpha, \lambda),\
\end{equation}
which in the small momentum limit leads to
\begin{equation}\label{shift_L}
F(\lambda) =- k \frac{\hat{L}_{\vartheta}(h ,\lambda)}{\vartheta(\lambda)} + \mathcal{\mathcal{O}}(k^2).
\end{equation}
The position of the particle and the hole can be expressed as functions of momentum and energy by solving the following equations
\begin{align}\label{smallk_kin1}
  p &= h + \frac{k}{k'(h)}  + \mathcal{O}(k^2) ,\\
  \frac{\omega}{k} &= v\left( h + \frac{k}{2 k'(h)} \right) + \mathcal{O}(k^2) ,\label{smallk_kin2}
\end{align}
with the dressed group velocity $v(\lambda)$ associated to the rapidity $\lambda$ defined by 
\begin{equation}\label{dressedvelocity}
v(\lambda) = \frac{d \varepsilon(\lambda)}{d k(\lambda)}.
\end{equation}
 {Given momentum $k$ and energy $\omega$, eq.~\eqref{smallk_kin2} is solved for the hole position $h$. Then position of particle $p$ is found from~\eqref{smallk_kin1}}.
The DSF at small momentum is given by by the single particle-hole contribution. Namely the single particle-hole contribution to the DSF 
fully gives the DSF up to (positive) corrections of order $k$, see Fig. \ref{fig:DSF} and section \ref{two_ph_and_more} 
\begin{equation}\label{saturation1ph}
S_{\hat{\rho}}(k,\omega) =  S^{1\text{ph}}_{\hat{\rho}}(k,\omega) \left( 1 + \mathcal{O}(k) \right).
\end{equation}
The single particle-hole contribution in the small momentum limit is given by
\begin{equation}\label{single_particle_DSF_smallk}
S^{1\text{ph}}_{ \hat{\rho}}(k,\omega) \simeq   (2\pi)^2\frac{ \rho(h ) \rho_h(p ) }{ |k v'(h)  k'(h)|}     \Big|\langle \vartheta| \hat{\rho}|\vartheta, h \to p  \rangle \Big|^2, 
\end{equation}
with $h,p$ fixed by conditions \eqref{smallk_kin1} and~\eqref{smallk_kin2}. The full single particle-hole form factors is given by 
\begin{align} \label{singlep_h_FF}
  |\langle \vartheta| \hat{\rho}(0) &|\vartheta, h \to p \rangle | =     2 \pi   \frac{  {\rho_t^{[p,h]}(h)  \rho_t^{[p,h]}(h)}{}}{\sqrt{\rho_t(p)\rho_t(h)   }       }   
     \frac{\pi \tilde{F}(p)}{\sin \pi\tilde{F}(p)} \frac{\sin \pi\tilde{F}(h)}{\pi \tilde{F}(h)}     \frac{\sin[ \pi \tilde{F}(h)]}{\vartheta(h) \sin [ \pi F(h)]}   \nn \\& \times
e^{ - \frac{c}{2} \pint_{-\infty}^{\infty} {\rm d}
    \lambda'  \frac{\vartheta(\lambda')F(\lambda') K(\lambda' - h)}{\lambda' - h} }\exp\left(\mathcal{B}(\vartheta, [p,h])\right)  \frac{{\rm Det}\Big(\boldsymbol{1} - K^{[p,h]}\Big)  }{ {\rm Det}\Big(\boldsymbol{1}- \hat{K}_{\vartheta} \Big)} ,
\end{align} 
and its limit $p \to h$ is remarkably simple 
\begin{equation}\label{singlephlimitk0}
  |\langle \vartheta| \hat{\rho} (0)|\vartheta, h \to p \rangle |  = k'(h) + \mathcal{O}(p-h).
\end{equation}
Notice that the denominator $|k v'(h)  k'(h)|$ in $S^{1\text{ph}}_{\hat{\rho}}(k,\omega)$ comes from the Jacobian of the change of variable $(p,h) \to (\omega,k)$ in the small $k$ limit.
Computing the static structure factor one finds the compressibility $\frac{\partial n}{\partial \mu}$ of the gas, which is the correct result for any finite temperature state
\begin{equation}
\lim_{k \to 0} \int_{-\infty}^{\infty} S_{\hat{\rho}}(k,\omega) \frac{{\rm d} \omega}{2 \pi } = \frac{\partial n}{\partial \mu}  ,
\end{equation}
with $\mu$ the chemical potential. At zero temperature {eq.~\eqref{single_particle_DSF_smallk} leads to }
\begin{equation}
\int_{-\infty}^{\infty} S_{\hat{\rho}}(k,\omega) \frac{{\rm d} \omega}{2 \pi } =   \frac{| k |}{v_s} + \mathcal{O}(k^2),
\end{equation}
with $v_s = v(q)$ and $q$  the Fermi momentum $k(q)= k_F = \pi n $. 

\subsection{Single particle-hole contribution and generalized detailed balance relation}
We here review the derivation of the detailed balance relation valid at small momentum $k$, as originally found in \cite{SciPostPhys.1.2.015}. 
We consider the exact single particle-hole DSF
\begin{equation}\label{single_particle_DSF}
S^{1\text{ph}}_{ \hat{\rho}}(k,\omega) = \Big(   \vartheta(h) (1-\vartheta(p))   \Big)    |\langle \vartheta| \hat{\rho}(0)|\vartheta, h \to p  \rangle  |^2  \ \frac{ k'(h ) k'(p ) }{| \det J_{p,h}|}  ,
\end{equation}
with position of particle and hole fixed by $\omega= \varepsilon(p) - \varepsilon(h)$ and $k= k(p)-k(h)$, with the form factors given in \eqref{singlep_h_FF} and the Jacobian of the change of variable $(p,h) \to (\omega,k)$
\begin{equation}\label{jacobian}
J_{p,h} = \begin{pmatrix}
 \varepsilon'(p) &  \varepsilon'(h) \\ k'(p) &k'(h)
\end{pmatrix}.
\end{equation}
We consider the ratio between the response of the system at positive and negative momenta and energies
\begin{equation}
 \frac{{S}_{\hat{\rho}}(-k,-\omega)}{ {S}_{\hat{\rho}}(k,\omega)} ,
\end{equation}
which for thermal states is known to be equal to $e^{\beta \omega}$. 
Since the excitation with particle-hole $(p,h)$ has energy $\omega = \varepsilon(p) - \varepsilon(h)$ and momentum $k = k(p) - k(h)$, the excitation with energy $-\omega$ and momentum $-k$ is the one with particle-hole given by $(h,p)$. Therefore the ratio of the the single particle-hole DSF is given by 
\begin{equation}\label{decomposition}
  \frac{S^{\text{1ph}}_{\hat{\rho}}(k,\omega)}{S^{\text{1ph}}_{\hat{\rho}}(-k,-\omega) } =    \frac{ \vartheta(h) \left( 1-\vartheta (p) \right) }{\vartheta(p) \left( 1-\vartheta (h) \right) }  \frac{\frac{   | \langle \vartheta | \hat{\rho} (0)| \vartheta, h \to p \rangle |^2    }{ | \det J_{p,h}  |}  }{ \frac{| \langle \vartheta | \hat{\rho}(0) | \vartheta, p \to h \rangle |^2    }{ | \det J_{h,p}  |}  }.
\end{equation}
The symmetry of the Jacobian implies that $|J_{p,h}| = | J_{h,p}|$. Numerically we observe that
\begin{equation}\label{eq:rationDB}
    \frac{   | \langle \vartheta | \hat{\rho}(0) | \vartheta, h \to p \rangle |^2    }  {    | \langle \vartheta | \hat{\rho}(0) | \vartheta, p \to h \rangle |^2    } = 1  ,
\end{equation}
which implies a particle-hole symmetry of the single particle-hole form factors.
It would be  desiderable to prove it analytically from the expression~\eqref{singlep_h_FF}, but we were not able to do so. 
Consider now a thermal state: using $\vartheta(\lambda) = (1+ e^{\beta (\varepsilon(\lambda)-\mu)})^{-1}$
we have 
\begin{equation}\label{eq:detailvartheta}
\frac{ \vartheta(h) \left( 1-\vartheta (p) \right) }{\vartheta(p) \left( 1-\vartheta (h) \right) }   = e^{\beta (\varepsilon(p) - \varepsilon(h))} = e^{\beta \omega}.
\end{equation}
We now consider non-thermal, parity invariant states, namely such that 
\begin{equation}
\int d \lambda \vartheta(\lambda) \lambda^{2n+1} = 0,
\end{equation}
for any integer $n$. 
For such states with $\vartheta(\lambda) = (1+ e^{ (\beta(\lambda)\varepsilon(\lambda)-\mu)})^{-1}$ we have
\begin{equation}
\frac{ \vartheta(h) \left( 1-\vartheta (p) \right) }{\vartheta(p) \left( 1-\vartheta (h) \right) } = e^{ (\varepsilon(p)\beta(p) - \varepsilon(h) \beta(h))} .
\end{equation}
In the small momentum limit with $p = h + k/k'(h)$ this last equation can be expanded in $k$ and we obtain\footnote{If the state is parity invariant, $\vartheta(\lambda) = \vartheta(-\lambda)$, then the correlation function obeys exactly $S_{\hat{\rho}}(k,\omega) = S_{\hat{\rho}}(-k,\omega)$. }
\begin{equation}\label{detailedbalance_smallk}
 \frac{ {S}^{1\text{ph}}_{\hat{\rho}}(k,\omega)}{  {S}^{1\text{ph}}_{\hat{\rho}}(-k,-\omega)}  = e^{\mathcal{F} (k, \omega) }    + \mathcal{O}(k^2),
\end{equation}
 where the function $ \mathcal{F} (k, \omega)$ {depends only on the state $|\vartheta \rangle$,  analogously to the thermal equilibrium case, and is given by}
\begin{equation}
\mathcal{F}(k,\omega) 
= k \frac{\partial \log (\vartheta^{-1} (h) -1)}{\partial k (h)} \Big|_{h = v^{-1}(\omega/k)},
\end{equation}
with the dressed velocity $v(h)$ given in equation \eqref{dressedvelocity}. While this relation is valid only for the single particle-hole contribution, using the result of equation \eqref{eq:detailedbalancesaved} we can show that at order $k$, namely on Euler hydrodynamic scales, the DSF satisfies a detailed balance relation for any parity invariant reference state~$ |\vartheta\rangle$
\begin{equation}
 \frac{ {S}_{\hat{\rho}}(k,\omega)}{  {S}_{\hat{\rho}}(-k,-\omega)}  =    e^{\mathcal{F}(k,\omega)} + \mathcal{O}(k^2),
\end{equation}
as originally found in \cite{SciPostPhys.1.2.015}. This form of generalized detailed balance allowed in \cite{SciPostPhys.3.3.023} to effectively ``measure" the distribution $\vartheta(\lambda)$ (and therefore its generalized temperatures $\beta(\lambda)$) by a measurement of $S_{\hat{\rho}}(k,\omega)$. { We expect the same form of detailed balance relation  for any operator that creates only pairs of particle-hole excitations, i.e. that conserves the total number of particles.}

For a thermal state instead detailed balance is exact at any value of momentum $k$, $S(k,\omega) = e^{\beta \omega} S(-k,-\omega)$. Indeed, from equation \eqref{eq:detailvartheta} and \eqref{eq:rationDB} we obtain that for any $k$ and thermal states~$| \vartheta \rangle$  
\begin{equation}
 \frac{ {S}^{1\text{ph}}_{\hat{\rho}}(k,\omega)}{  {S}^{1\text{ph}}_{\hat{\rho}}(-k,-\omega)}  = e^{\beta \omega}.
\end{equation}
This implies that when $| \vartheta \rangle $ is a thermal state each $m$-th particle-hole contribution satisfies independently the detailed balance relation: $S^{m\text{ph}}_{\hat{\rho}}(k,\omega) = S^{m\text{ph}}_{\hat{\rho}}(-k,-\omega) e^{\beta \omega}$.   
%

\subsection{Single particle-hole contribution and Generalized Hydrodynamics}
We show here that  what was found in \cite{SciPostPhys.1.2.015}, namely that the small momentum limit of the density DSF is given by the single particle-hole contribution
\begin{equation}\label{single_particle_DSF_2}
\underset{\frac{\omega}{k} = \kappa }{ \lim_{k,\omega \to 0}} S_{\hat{\rho}}(k,\omega) =\underset{\frac{\omega}{k} = \kappa }{ \lim_{k,\omega \to 0}}  S^{1\text{ph}}_{\hat{\rho}}(k,\omega),
\end{equation}
and that its small momentum limit is given in terms of the inverse of the dressed velocity $v(h)$ \eqref{dressedvelocity}
\begin{equation}\label{single_particle_DSF_2_limit}
  \underset{\frac{\omega}{k} = \kappa }{ \lim_{k,\omega \to 0}} S^{1\text{ph}}_{\hat{\rho}}(k,\omega) = (2\pi)^2 \frac{\rho(h) \rho_h(h)}{ | k v'(h) k'(h)| } \left( k'(h) \right)^2   \Big|_{h =  v^{-1}(\kappa)}  ,
\end{equation}
is compatible with the predictions of Generalized Hydrodynamics (GHD) \cite{arXiv.1605.09790,Ben}. In the context of GHD it was shown \cite{DoynSpohn,1711.04568} that given the density $\hat{q}(x)$ of a generic conserved operator $\hat{Q} = \int \hat{q}(x) dx$, such that $[H,\hat{Q} ]=0$, the hydrodynamic description of the excitations implies a generic form for the asymptotic correlations 
\begin{equation}
\langle \hat{q}(x,t)  \hat{q}(0,0) \rangle \simeq  (2\pi)\int {\rm d}h  \delta(x - v(h) t )  \rho(h) (1- \vartheta(h)) ( q^{\text{dr}}(h))^2,
\end{equation}
at large $x$ and $t$ with $x/t$ fixed (the so-called Euler scale) and 
with the dressed single-particle eigenvalue of the charge given by
\begin{equation}
q^{\text{dr}}(h) =q(h) + \int_{-\infty}^{\infty} {\rm d}\alpha \hat{L}_{\vartheta}(h,\alpha) q(\alpha).
\end{equation}
with $\hat{L}_{\vartheta}(\lambda,\mu)$ the resolvent \eqref{L_int_eq} and $q(\lambda)$ the eigenvalue of the charge $\hat{Q}$ on a single particle state $| \lambda \rangle$. Going to Fourier space this result implies the following form for the DSF  at small momentum and energy
\begin{equation}
S_{\hat{q}}(k,\omega) \simeq (2\pi) \int {\rm d}h \delta(\omega - v(h) k)  \rho(h) (1- \vartheta(h)) ( q^{\text{dr}}(h))^2 = (2\pi) \frac{ \rho(h) (1- \vartheta(h)) ( q^{\text{dr}}(h))^2}{| k v'(h) |},
\end{equation}
which is in accord with our result for the density DSF \eqref{single_particle_DSF_2_limit} after using that
\begin{equation}
k'(h) = n^{\text{dr}} = 2 \pi \rho_t(h)  \quad \text{ with }  \quad n=1.
\end{equation}
GHD therefore implies that the result \eqref{single_particle_DSF_2_limit} and \eqref{single_particle_DSF_2} also applies to the DSF of any globally conserved operator $ S_{\hat{q}}(k,\omega)$. Namely that in the small momentum limit this is saturated by the single particle-hole contribution with amplitude given by
\begin{equation}
\underset{\frac{\omega}{k} = \kappa }{ \lim_{k,\omega \to 0}}S_{\hat{q}}(k,\omega) =  (2\pi)^2\frac{ \rho(h ) \rho_h(h ) }{ |k v'(h)  k'(h)|}  \Big[  \lim_{p \to h}   \Big|\langle \vartheta| \hat{q}|\vartheta, h \to p  \rangle \Big|^2  \Big]   \Big|_{h =v^{-1}( \kappa)} ,
\end{equation}
with the following universal form for the form factors in the small momentum limit
\begin{equation}
 \lim_{p \to h} | \langle \vartheta| \hat{q}|\vartheta, h \to p  \rangle | = q^{\text{dr}}(h) .
\end{equation}
However GHD does not provide the corrections $\mathcal{O}(p-h)$ and in general it only incorporates the leading term \eqref{singlephlimitk0}, in the small momentum limit, of the full single particle-hole form factor \eqref{singlep_h_FF}. Up to now the only method to get the full form factor expression is by taking the thermodynamic limit of the finite size expressions as we do here for the density operator $ \hat{\rho}(x)$. 

%
%
%

\section{Two and more particle-hole contributions} \label{two_ph_and_more}
In this section we analyze higher order (in number of particle-hole pairs) contributions to the DSF.
The analysis at small momentum shows that the DSF is organized in the particle-hole contributions
\begin{equation}\label{single_particle_DSF_3}
S_{\hat{\rho}}(k,\omega) =  \sum_{m \geq 1}S^{m\text{ph}}_{\hat{\rho}}(k,\omega),
\end{equation}
where each contribution at small $k$ is of order 
\begin{equation}
S^{m\text{ph}}_{\hat{\rho}}(k,\omega) \sim \mathcal{O}(k^{m-2}). 
\end{equation}
{We will show that this is the case first for the 2 particle-hole contribution. The generalization to arbitrary number of particle-hole pairs will then follow immediately.} 
According to eq.~\eqref{mph_contribution} this contribution is given by
\begin{align}
 S^{2\text{ph}}_{\hat{\rho}}(k,\omega) & = \frac{1}{4}  \int {\rm d}h_1\, \rho(h_1) \fint {\rm d}p_1 \rho_h(p_1 )  \int {\rm d}h_2\, \rho(h_2) \fint {\rm d}p_2 \rho_h(p_2)   |\langle \vartheta| \hat{\rho}(0)|\vartheta, \mathbf{h} \to \mathbf{p} \rangle |^2  \nn \\
& \times   \delta(\omega - \varepsilon(p_1) -  \varepsilon(p_2) + \varepsilon(h_1)  +\varepsilon(h_2)   )
\delta(k - k(p_1) -  k(p_2) + k(h_1)  +k(h_2)   ).
\end{align}
We are not interested in precise evaluation of this formula, but just in establishing the leading order $k$. 
To this end we change variables in the integrals, from $h_1,p_1$ to the variables $\varepsilon_1 = \varepsilon(p_1) - \varepsilon(h_1)$, $k_1 = k(p_1) - k(h_1)$. We do the same for $h_2,p_2$ with $\varepsilon_2,k_2$. We obtain
\begin{align}\label{2phint1}
 S^{2\text{ph}}_{\hat{\rho}}(k,\omega)& = \frac{1}{4} \fint {\rm d}k_1 {\rm d}k_2  \int {\rm d} \varepsilon_1 {\rm d} \varepsilon_2\, \rho(h_1)   \rho_h(p_1 ) \, \rho(h_2)  \rho_h(p_2)\nn \\&   \frac{|\langle \vartheta| \hat{\rho}(0)|\vartheta, \mathbf{h} \to \mathbf{p} \rangle |^2}{|  \det J_{p_1,h_1} \det J_{p_2,h_2}|}   \delta(\omega - \varepsilon_1 - \varepsilon_2)
\delta(k -k_1 - k_2   ),
\end{align}
with position of holes and particles expressed in terms of $\varepsilon_i,k_i$ and the Jacobian as in~\eqref{jacobian}. The integrations of over $k_2$ and $\epsilon_2$ can be performed what fix their values by the momentum energy conservation
\begin{align}\label{2phint1_final} 
 S^{2\text{ph}}_{\hat{\rho}}(k,\omega) = \frac{1}{4} \fint {\rm d}k_1 \int {\rm d} \varepsilon_1 \, \rho(h_1)   \rho_h(p_1 )  \, \rho(h_2)  \rho_h(p_2)    \frac{|\langle \vartheta| \hat{\rho}(0)|\vartheta, \mathbf{h} \to \mathbf{p} \rangle |^2}{|  \det J_{p_1,h_1} \det J_{p_2,h_2}|}.
\end{align}
We now restrict to the regime where $k$ is small and $\omega/k$ is finite. In this limit $k_1$ and $k_2=(k-k_1)$ are both small. Therefore the particle-hole pairs corresponding to $k_1$ and $k_2$ are also small. In the next subsection we show that the form factors in this limit has a leading part of order $k^0$. The integrals are both of order $k$ (the energy is, in this limit, a linear function of $k$) and the Jacobians are also of order $k$. Therefore~\eqref{2phint1_final} is of order~$k^0$ and it gives a subleading term to the single particle-hole contribution $S^{1\text{ph}}_{\hat{\rho}}(k,\omega)$ which is of order $k^{-1}$ \footnote{Notice that this implies that static structure factor also has a linear coefficient in $k$ such as $\int S(k,\omega) \frac{d\omega}{2\pi} = S(0) + \mathcal{O}(k) + \mathcal{O}(k^2)$. On the other hand usually, for small temperatures at least, the terms proportional to $k$ as such as the ones coming from \eqref{2phint1_final} are small.}. Extending the logic to $m$ particle-hole is straightforward. The form factors, for arbitrary number of particle-hole excitations, is always of order $k^0$. Each integration over a pair of particle-hole can be converted into corresponding energy and momentum conservations divided by the Jacobian. This bring a factor $k$ for each particle-hole pair and the momentum and energy conservation eliminate two of the integrals, giving the final order $k^{m-2}$. The same logic is expected to be applicable to the DSF of any particle-conserving operator $\hat{q}$. Such an operator indeed can only create particle-hole pairs and therefore the same arguments apply, provided that its form factors are of order $k^0$.

Notice that even if the leading two particle-hole contribution \eqref{2phint1_final} is of order $1$, this does not spoil the detailed balance in the linear order in $k$, equation \eqref{detailedbalance_smallk}. This is because $S^{2\text{ph}}_{\hat{\rho}}(k,\omega)$ is an even function of $\omega$ for symmetric states
(This is because, in the leading order in $k$ the 2 particle-hole form factors has a particle-hole symmetry.)
Therefore taking $\tilde{S}^{1\text{ph}}_{\hat{\rho}}(k,\omega) = |k| {S}^{1\text{ph}}_{\hat{\rho}}(k,\omega) \sim \mathcal{O}(k^0) $ and expanding the ratio of the DSF computed with opposite energies we obtain
\begin{align}\label{eq:detailedbalancesaved}
  \frac{S_{\hat{\rho}}(k,\omega)}{S_{\hat{\rho}}(k,-\omega)} &= \frac{\tilde{S}^{1\text{ph}}_{\hat{\rho}}(k,\omega) +  |k|  S^{2\text{ph}}_{\hat{\rho}}(k,\omega)}{ \tilde{S}^{1\text{ph}}_{\hat{\rho}}(-k,-\omega) +  |k| S^{2\text{ph}}_{\hat{\rho}}(k,\omega)}  + \mathcal{O}(k^2) \nn \\
   &= \frac{S^{1\text{ph}}_{\hat{\rho}}(k,\omega)  }{S^{1\text{ph}}_{\hat{\rho}}(-k,-\omega) } -| k|  S^{2\text{ph}}_{\hat{\rho}}(k,\omega) \frac{(S^{1\text{ph}}_{\hat{\rho}}(k,\omega)  - S^{1\text{ph}}_{\hat{\rho}}(-k,-\omega) )}{\left(\tilde{S}^{1\text{ph}}_{\hat{\rho}}(-k,-\omega)\right)^2 } + \mathcal{O}(k^2).
\end{align}
Since $(S^{1\text{ph}}_{\hat{\rho}}(k,\omega)  - S^{1\text{ph}}_{\hat{\rho}}(-k,-\omega) ) \sim \mathcal{O}(k)$ the second term is also of order $k^2$, leading to the detailed balance expression \eqref{detailedbalance_smallk}
 \begin{align}
& \frac{S_{\hat{\rho}}(k,\omega)}{S_{\hat{\rho}}(-k,-\omega)} =  \frac{S^{1\text{ph}}_{\hat{\rho}}(k,\omega)  }{S^{1\text{ph}}_{\hat{\rho}}(-k,-\omega) } + \mathcal{O}(k^2) = e^{\mathcal{F}(k,\omega)} + \mathcal{O}(k^2).
\end{align}

 \subsection{Two particle-hole contribution in the small momentum limit}\label{exactcompFF}
 We here show that in the small momentum limit, with $p_1 = h_1 + k_1/k'(h_1)$ and $p_2 = h_ 2 +  (1-k_1)/k'(h_2)$ and $k_1  \to k \kappa $ with $k \to 0$, we have a well defined form factor as function of $h_1,h_2$ and $\kappa$ analogously to the single particle-hole case \eqref{singlephlimitk0}. 
For such excitations the shift function is given simply by the sum of the shift function for each excitation
\begin{equation}\label{backflow_resolvent_3}
 F(\lambda) = - \frac{1}{\vartheta(\lambda)} \left( (p_1 - h_1) \hat{L}_{\vartheta}(h_1,\lambda)  +  (p_2 - h_2) \hat{L}_{\vartheta}(h_2,\lambda) \right) + \mathcal{O}(k^2 ) .\
\end{equation}
In the form factors the only relevant piece is the matrices $A_{ij}$ and $B_{ij}$, as almost all the others are close to one (except few terms from $\mathcal{A}(\vartheta, \phset)$), as $F(\lambda) \sim k$ in the small momentum limit. We have 
\begin{equation}
\resatilde^{[\mathbf{p}, \mathbf{h}]} \xrightarrow{\mathbf{p} \rightarrow \mathbf{h}} -\frac{(p_1 - h_i)( p_2 - h_i)}{(h_j  - h_i)_{j\neq i}} \frac{\vartheta(h_i)}{2 \pi}, \qquad i,j = 1,2,
\end{equation}
and 
\begin{equation}
\left[ \lim_{\lambda \to h_j} \frac{L^{[\mathbf{p}, \mathbf{h}]}(h_i,\lambda)}{\tilde{a}^{[\mathbf{p}, \mathbf{h}]}(\lambda)} \right]   \xrightarrow{\mathbf{p} \rightarrow \mathbf{h}} \frac{2 \pi \hat{L}_{\vartheta}(h_i,h_j)}{\vartheta(h_j)}.
\end{equation}
Neglecting corrections of order $k$ (namely we set $\frac{ p_1-p_2}{h_1- h_2}  =1$ and $\frac{p_1- h_2}{h_1- h_2} = 1$) we obtain for the matrices $A_{ij}$ and $B_{ij}$
\begin{equation}
A_{ij} = \begin{pmatrix}
 1 -  \frac{(p_1- h_1) \hat{L}_{\vartheta}(h_1,h_1)}{  (p_1 - h_1) \hat{L}_{\vartheta}(h_1,h_1) + (p_2- h_2) \hat{L}_{\vartheta}(h_2,h_1) }  & -\frac{\vartheta(h_1)}{\vartheta (h_2)} \frac{(p_1- h_1)  \hat{L}_{\vartheta}(h_1,h_2)}{ (p_1 - h_1) \hat{L}_{\vartheta}(h_1,h_1) + (p_2- h_2) \hat{L}_{\vartheta}(h_2,h_1) } \\
  -\frac{\vartheta(h_2)}{\vartheta (h_1)}   \frac{ (p_2 - h_2) \hat{L}_{\vartheta}(h_2,h_1)}{  (p_1 - h_1) \hat{L}_{\vartheta}(h_1,h_2) + (p_2- h_2) \hat{L}_{\vartheta}(h_2,h_2)}  & 1 - \frac{ (p_2 - h_2) \hat{L}_{\vartheta}(h_2,h_2)}{    (p_1 - h_1) \hat{L}_{\vartheta}(h_1,h_2) + (p_2- h_2) \hat{L}_{\vartheta}(h_2,h_2)}
   \end{pmatrix},
 \end{equation}
\begin{equation}
  B_{ij} =  \begin{pmatrix}
    \frac{(p_1- h_1) \rho(h_1) (2 \pi )\rho_t(h_1)}{  (p_1 - h_1) \hat{L}_{\vartheta}(h_1,h_1) + (p_2- h_2) \hat{L}_{\vartheta}(h_2,h_1) }  & \frac{(p_1- h_1)  \rho(h_1) (2 \pi )\rho_t(h_2)}{ (p_1 - h_1) \hat{L}_{\vartheta}(h_1,h_1) + (p_2- h_2) \hat{L}_{\vartheta}(h_2,h_1) } \\
    \frac{ (p_2 - h_2) \rho(h_2) (2 \pi )\rho_t(h_1)}{  (p_1 - h_1) \hat{L}_{\vartheta}(h_1,h_2) + (p_2- h_2) \hat{L}_{\vartheta}(h_2,h_2) }  & \frac{ (p_2 - h_2) \rho(h_2) (2 \pi )\rho_t(h_2)}{    (p_1 - h_1) \hat{L}_{\vartheta}(h_1,h_2) + (p_2- h_2) \hat{L}_{\vartheta}(h_2,h_2)}
  \end{pmatrix}.
\end{equation}
The determinant of $A_{ij}$ is zero as it should, see \eqref{detAzero} and Appendix \ref{sec:constant_alpha}.  The full form factors in the leading order in $k$, which means the leading order in $p_j-h_j$, then reads 
\begin{align}\label{2phFF_smallk}
 |\langle \vartheta|   & \hat{\rho}(0)|\vartheta, \{h_1,h_2\}\to \{p_1,p_2\} \rangle | =   \prod_{k=1}^2 \frac{F(h_k)}{\rho_t(h_k)(p_k- h_k)} {\rm det}\left(A_{ij} +B_{ij}\right) + \mathcal{O}( k_1) \nn \\&
 =    \Big(  { \frac{k_1}{k'(h_1) } 2 \pi \rho_t(h_1)   + \frac{(1-k_1)}{k'(h_2)} 2 \pi \rho_t(h_2)  }{ }\Big) \nn  \\& \times \left(  \frac{ \hat{L}_{\vartheta}(h_1,h_2)}{\rho(h_2)} \frac{ k'(h_2) }{1-k_1 } +  \frac{    \hat{L}_{\vartheta}(h_2,h_1)}{\rho(h_1)} \frac{ k'(h_1) }{k_1 } \right)  + \mathcal{O}(k_1).
\end{align}
By rescaling $k_1  \to \kappa k$ and taking the limit $k \to 0$ it is easy to see that the whole form factor is of order $k^0$. Increasing number of particle-hole pairs simply extends the product and the matrices. The structure is however the same and  form factors for any number of particle-hole pairs in the small momentum limit is of order $k^0$. Notice  that for large values of the coupling $c$, the resolvent vanishes as $\hat{L}_{\vartheta}(h_1,h_2) \sim 1/c$ and therefore the two-particle hole contribution $S^{2\text{ph}}_{\hat{\rho}}$ decays as $1/c^2$ as expected.  
%

\section{Dressing of particle-hole excitations at zero temperature}\label{sec:dressing}

In this section we study the behavior of the form factors when the thermodynamic state $| \vartheta\rangle$ represents a critical state. The discontinuities in the filling function $\vartheta(\lambda)$ affect the structure of the thermodynamic form factors and excitations created in the vicinity of the discontinuities lead to divergences. A priori this is not a surprise. The thermodynamic limit of the form factors was derived in~\cite{Smooth_us} under an assumption that the filling function is smooth and therefore using it to study correlation functions of critical states seems problematic. However, in~\cite{SciPostPhys.1.2.015} we have shown that computing the small momentum limit of the ground state static structure factor leads to a correct answer. In this section we show that in general we can extract small momentum information about the critical states from the form factors.


To show the difference between the critical and non-critical states we will consider form factors with an excited state consisting of one dominant excitation, carrying most of the momentum and energy of the excited state, and many small excitations, namely with vanishing energy and momentum, that dress it (\textit{soft modes}). In the two following sections we consider what happens when the state is non-critical (smooth filling function) and when the state is critical (discontinuous filling function). In the first case the contribution from these form factors has zero measure and no dressing is needed for each particle-hole excitation. In the second case, the form factors can be divergent when the soft modes are localized in the vicinity of the discontinuities of the filling function, i.e. the Fermi momenta. We show that this implies that for a single particle-hole excitation with momentum $k \geq k^*$ its form factors must be dressed with soft modes excitations whose contribution is not negligible. We denote this as the dressing threshold $k^{*}$ 
\begin{equation}\label{dressing_th}
k^* = \Big|  \frac{\sqrt{K}}{2\partial_q F(q,q)} \Big|,
\end{equation}  
with $q$ the Fermi momentum and $K$ the Luttinger liquid parameter $2 \pi \rho_t(q) = \sqrt{K}$. In the next session \ref{sec:edge} will also show that in the small momentum limit the DSF close to the two edges (in correspondence with the Lieb I and II dispersion relations $\epsilon_1(k),\epsilon_2(k)$) we have
\begin{equation}
S(k,\omega) \sim \Big|\omega - \epsilon_{1,2}(k)\Big|^{\mp k /k^{*}}.
\end{equation} 
Therefore when $k \geq  k^{*}$ the DSF $S(k,\omega)$ displays a non-integrable singularity at $\omega = \epsilon_1$, which signals a divergence of the contributions given by the soft modes to the form factors.


We consider one single dominant particle-hole excitation and $m$ soft modes. Recall that each particle-hole excitation contributes $k(p_i, h_i) = k(p_i)- k(h_i)$ and $\varepsilon(p_i, h_i) = \varepsilon(p_i) - \varepsilon(h_i)$ to the momentum and the energy. We assume that $\{p_0,h_0\}$ pair is dominant, that is
\begin{equation}
  \frac{k - k(p_0,h_0)}{k} \ll 1, \qquad \frac{\omega - \varepsilon(p_0,h_0)}{\omega} \ll 1.
\end{equation}
In other words, for finite $k$ and $\omega$ we have
\begin{equation}
  \frac{1}{k}\sum_{i=1}^m k(p_i,h_i) \ll 1, \qquad \frac{1}{\omega}\sum_{i=1}^m \varepsilon(p_i,h_i) \ll 1.
\end{equation}
The momentum and energy of the particle hole pair are, c.f. eqs.~\eqref{k} and~\eqref{omega},
\begin{align}
  k(p_i,h_i) &= p_i - h_i - \int_{-\infty}^{\infty} {\rm d}\lambda\, \vartheta(\lambda) F(\lambda|p_i,h_i), \\
  \varepsilon(p_i,h_i) &= p_i^2 - h_i^2 - 2\int_{-\infty}^{\infty} {\rm d}\lambda\, \lambda\vartheta(\lambda) F(\lambda|p_i,h_i).
\end{align}
For the left hand side to be small, the difference $p_i-h_i$ must be small. Then the back-flow simplifies
\begin{equation}
  F(\lambda|p_i, h_i) = - (p_i - h_i)\frac{\hat{L}_{\vartheta}(h_i,\lambda)}{\vartheta(\lambda)} + \mathcal{O}((p_i-h_i)^2).
\end{equation}
For the ground state these small excitations are only possible in the vicinity of the Fermi edges $\pm q$. For a generic state the small excitations can be created along the whole distribution of rapidities. 
The total back flow is
\begin{equation}
  F(\lambda|\{p_i,h_i\}) = F(\lambda|p_0,h_0) - \sum_{i=1}^m (p_i-h_i) \frac{\hat{L}_{\vartheta}(h_i, \lambda)}{\vartheta(\lambda)},
\end{equation}
and is dominated by the back-flow of the dominating excitation. Let us write
\begin{equation}
  F_0(\lambda) = F(\lambda|p_0,h_0).
\end{equation}

We compute now the leading part of the form-factor in the particle-hole difference $p_i - h_i$ of the small excitations. The result is 
\begin{align}
  |\langle \vartheta| \hat{\rho}(0)|&\vartheta, \{h_j\rightarrow p_j\}_{j=0}^m\rangle| = |\langle \vartheta| \hat{\rho}(0)|\vartheta, \{h_o\rightarrow p_o\}\rangle | \left[\prod_{k=1}^m \frac{F_0(h_k)}{\rho_t(h_k)} \frac{\vartheta(p_k) \sin \pi \vartheta(h_k) F_0(h_k)}{\vartheta(h_k) \sin \pi \vartheta(p_k)F_0(p_k)} \right] \nn\\
  &\times \frac{\prod_{i<j=1}^m (h_i - h_j)(p_i-p_j)}{\prod_{i,j=1}^m(p_i - h_j)} \exp\left( \sum_{k=1}^m \pint_{-\infty}^{+\infty} {\rm d}\lambda \frac{\tilde{F}(\lambda)(h_k-p_k)}{(\lambda-h_k)(\lambda-p_k)} \right), \label{ff_dressing}
\end{align}
and its derivation is given in Appendix~\ref{derivation_dressing}.

\subsection{Non-critical states: smooth filling functions}
We analyze now the contribution of the form-factors with a single dominant excitation to the correlation function of a non-critical state. The filling function is smooth and from eq.~~\eqref{ff_dressing} we get
\begin{equation}
  |\langle \vartheta| \hat{\rho}(0)|\vartheta, \phset\rangle| = |\langle \vartheta| \hat{\rho}(0)|\vartheta, p_0,h_0\rangle | \left[\prod_{k=1}^m \frac{F_0(h_k)}{\rho_t(h_k)}\right] \frac{\prod_{i<j=1}^m (h_i - h_j)(p_i-p_j)}{\prod_{i,j=1}^m(p_i - h_j)}.
\end{equation}
Consider the $(m+1)$ particle hole contribution to the correlation function
\begin{equation}
  A_m = \fint {\rm d}{\mathbf p}_{m+1} {\rm d}{\mathbf h}_{m+1} |\langle \vartheta| \hat{\rho}(0)|\vartheta, \phset\rangle|^2 \delta(k - k(\mathbf{p}, \mathbf{h}))\delta(\omega - \varepsilon(\mathbf{p}, \mathbf{h})),
\end{equation}
where we use the finite part integral. Restricting the integral over the $m$ particle-hole pairs to small excitations we obtain
\begin{equation}
A_m = \fint {\rm d}p_0 {\rm d}h_0 |\langle \vartheta| \hat{\rho}(0)|\vartheta, \phset\rangle|^2 B_m(k,\omega, p_0, h_0) \delta(k - k_0)\delta(\omega - \omega_0),
\end{equation}
where
\begin{equation}
  B_m  = \fint {\rm d}{\mathbf p}_{m} {\rm d}{\mathbf h}_{m}  \left[\prod_{k=1}^m \frac{F_0(h_k)}{\rho_t(h_k)} \frac{\prod_{i<j=1}^m (h_i - h_j)(p_i-p_j)}{\prod_{i,j=1}^m(p_i - h_j)} \right]^2. 
\end{equation}
The integral in $B_m$ is over regions where $p_i - h_i$ is small so that the approximation to the form factors can be used. We also used that the momentum and energy is determined mainly by the dominant excitation to simplify the Dirac's $\delta$-function. The corrections are proportional to $(p_i-h_i)$ and can be neglected. The integrand in $B_m$ has a double pole whenever positions of particle and hole coincide. These double poles are regularized by the finite part integral according to the prescription~\eqref{finite_part_integral}. Therefore $B_m$ is finite. Each finite part integral is over a small region of holes around the position of particles. This is the phase space given to soft modes and $B_m$ is roughly proportional to the volume of the phase space that we allow for them. This implies that if we consider a single particle-hole excitation and we dress it with soft modes, their contribution vanishes in the limit of a vanishing phase space. Therefore there is no dressing of the form factors for non-critical states. The situation is different if the state is critical, as the form factors has poles when the small particle-hole excitations are taken close to the edges of the Fermi sea.


\subsection{Critical states: dressing threshold \texorpdfstring{$k^{*}$}{k*}}

Let us consider the archetypical critical state, the ground state. For the ground state the filling function $\vartheta(\lambda)$ equals to $1$ for $\lambda \in [-q,q]$ and zero otherwise. Therefore particles must have $|p_j| > q$ while holes $|h_j| < q$. For the form factors we get 
\begin{align}
  |\langle \vartheta| \hat{\rho}(0)|&\vartheta, \mathbf{h} \to \mathbf{p}\rangle| = |\langle \vartheta| \hat{\rho}(0)|\vartheta, p_0,h_0\rangle | \left[\prod_{k=1}^m \frac{\sin \pi F(h_k)}{\pi \rho_t(h_k)}  \right] \frac{\prod_{i<j=1}^m (h_i - h_j)(p_i-p_j)}{\prod_{i,j=1}^m(p_i - h_j)} \nn\\
  &\times \exp\left( \sum_{k=1}^m  \pint_{-\infty}^{\infty} {\rm d}\lambda \vartheta(\lambda)\frac{F_0(\lambda)(h_k-p_k)}{(\lambda-h_k)(\lambda-p_k)} \right).
\end{align}
Let us focus on a single excitation and analyse the structure of the principal value integral. We seperate the particle and hole parts
\ea {
  \frac{(h_k-p_k)}{(\lambda-h_k)(\lambda-p_k)} = \frac{1}{\lambda-h_k} - \frac{1}{\lambda-p_k},
}
and consider first the hole contribution. We have
\ea {
  \pint_{-\infty}^{\infty} {\rm d}\lambda\, \frac{\vartheta(\lambda) F_0(\lambda)}{\lambda-h_k} &= \lim_{\epsilon\rightarrow 0} \left(\int_{-\infty}^{h_k-\epsilon} {\rm d}\lambda\, \frac{\vartheta(\lambda) F_0(\lambda)}{\lambda-h_k} + \int_{h_k+\epsilon}^{\infty} {\rm d}\lambda\, \frac{\vartheta(\lambda) F_0(\lambda)}{\lambda-h_k} \right) \nonumber \\
  &= \lim_{\epsilon\rightarrow 0} \left(\int_{-q}^{h_k-\epsilon} {\rm d}\lambda\, \frac{F_0(\lambda)}{\lambda-h_k} + \int_{h_k+\epsilon}^{q} {\rm d}\lambda\, \frac{F_0(\lambda)}{\lambda-h_k} \right),
}
where in the first line we used the definition of the principal value integral and in the second line we used that the filling function vanishes beyond the interval $[-q,q]$ and is $1$ within.
For the excitation to be small, position of the hole must be close to one of the  edges. Let us assume that $h_k \sim q$. Then the second integral can be simplified, making an error of order~$q -h_k$, in the following way
\ea {
  \pint_{-\infty}^{\infty} {\rm d}\lambda\, \frac{\vartheta(\lambda) F_0(\lambda)}{\lambda-h_k} &= \lim_{\epsilon\rightarrow 0} \left(\int_{-q}^{h_k-\epsilon} {\rm d}\lambda\, \frac{F_0(\lambda)}{\lambda-h_k} + F_0(q) \log \frac{q - h_k}{\epsilon} \right) \nonumber \\
 &= \lim_{\epsilon\rightarrow 0} \left(\int_{-q}^{h_k-\epsilon} {\rm d}\lambda\, \frac{F_0(\lambda)}{\lambda-h_k} - F_0(q) \log \epsilon \right) + F_0(q) \log (q - h_k).
}
The first expression is smooth as a function of $h_k$ thus we can set $h_k = q$. Therefore for a hole in the vicinity of $q$ we get
\ea {
  \pint_{-\infty}^{\infty} {\rm d}\lambda\, \frac{\vartheta(\lambda) F_0(\lambda)}{\lambda-h_k} = \ppint_{-q}^{q} {\rm d}\lambda\, \frac{F_0(\lambda)}{\lambda-q} + F_0(q) \log (q - h_k),
}
where we defined a version of the principal value integral with the pole at the boundary
\ea {
  \ppint_a^b {\rm d}\lambda \frac{f(\lambda)}{\lambda - b} = \lim_{\epsilon\rightarrow 0} \left(\int_{a}^{b-\epsilon} {\rm d}\lambda\, \frac{f(\lambda)}{\lambda-b} - f(b) \log \epsilon \right).
}
Similar situation happens for the particle part and we find
\ea {
  \pint_{-\infty}^{\infty} {\rm d}\lambda\, \frac{\vartheta(\lambda) F_0(\lambda)}{\lambda-p_k} = \ppint_{-q}^{q} {\rm d}\lambda\, \frac{F_0(\lambda)}{\lambda-q} + F_0(q)\log(q-p_k).
}
The whole principal value integral becomes
\ea {
  \pint_{-\infty}^{\infty} {\rm d}\lambda \vartheta(\lambda)\frac{F_0(\lambda)(h_k-p_k)}{(\lambda-h_k)(\lambda-p_k)} = \ppint_{-q}^{q} {\rm d}\lambda \frac{F_0(\lambda)(h_k-p_k)}{(\lambda-h_k)(\lambda-p_k)} + F_0(q) \log\frac{q - h_k}{q - p_k}.
}
The contribution of the $P_+$ integral is bounded and proportional to $p_k-h_k$. Therefore it can be neglected. The leading part of the form factors is
\ea {
  |\langle \vartheta| \hat{\rho}(0)|\vartheta, \phset\rangle| =& |\langle \vartheta| \hat{\rho}(0)|\vartheta, p_0,h_0\rangle | \left[\prod_{k=1}^m \frac{\sin \pi F(h_k)}{\pi \rho_t(h_k)}  \right] \prod_{k=1}^m \left(\frac{q-h_k}{q-p_k}\right)^{F_0(q)} \nonumber \\
  & \times  \frac{\prod_{i<j=1}^m (h_i - h_j)(p_i-p_j)}{\prod_{i,j=1}^m(p_i - h_j)}.
}
Following in the same way as for the smooth distribution function we find
\ea {
 B_m  = \prod_{k=1}^m \fint {\rm d}p_{k} {\rm d}h_{k} \left[\frac{\sin \pi F(h_k)}{\pi \rho_t(h_k)} \left(\frac{q-h_k}{q-p_k}\right)^{F_0(q)} \frac{\prod_{i<j=1}^m (h_i - h_j)(p_i-p_j)}{\prod_{i,j=1}^m(p_i - h_j)} \right]^2 . 
}
The finite part integral regularizes the pole at $p_i-h_j$. However the divergence for $p_k\sim q$ remains and therefore the integral cannot be evaluated. The cure would be to regularize the remaining integral by going back to a finite system. We would then find that the contribution to $B_m$ is finite but does not scale with the size of the region. Moreover, the distance $q-p_k$ is resolved in finite system as
\ea {
  q - p_k \sim \frac{1}{L},
}
and leads to a contribution to a correlation functions that has a fractional power in the system size. Therefore to obtain a finite correlation function one would need to sum over all possible small excitations.

\begin{figure}
  \center
  \includegraphics[scale=0.5]{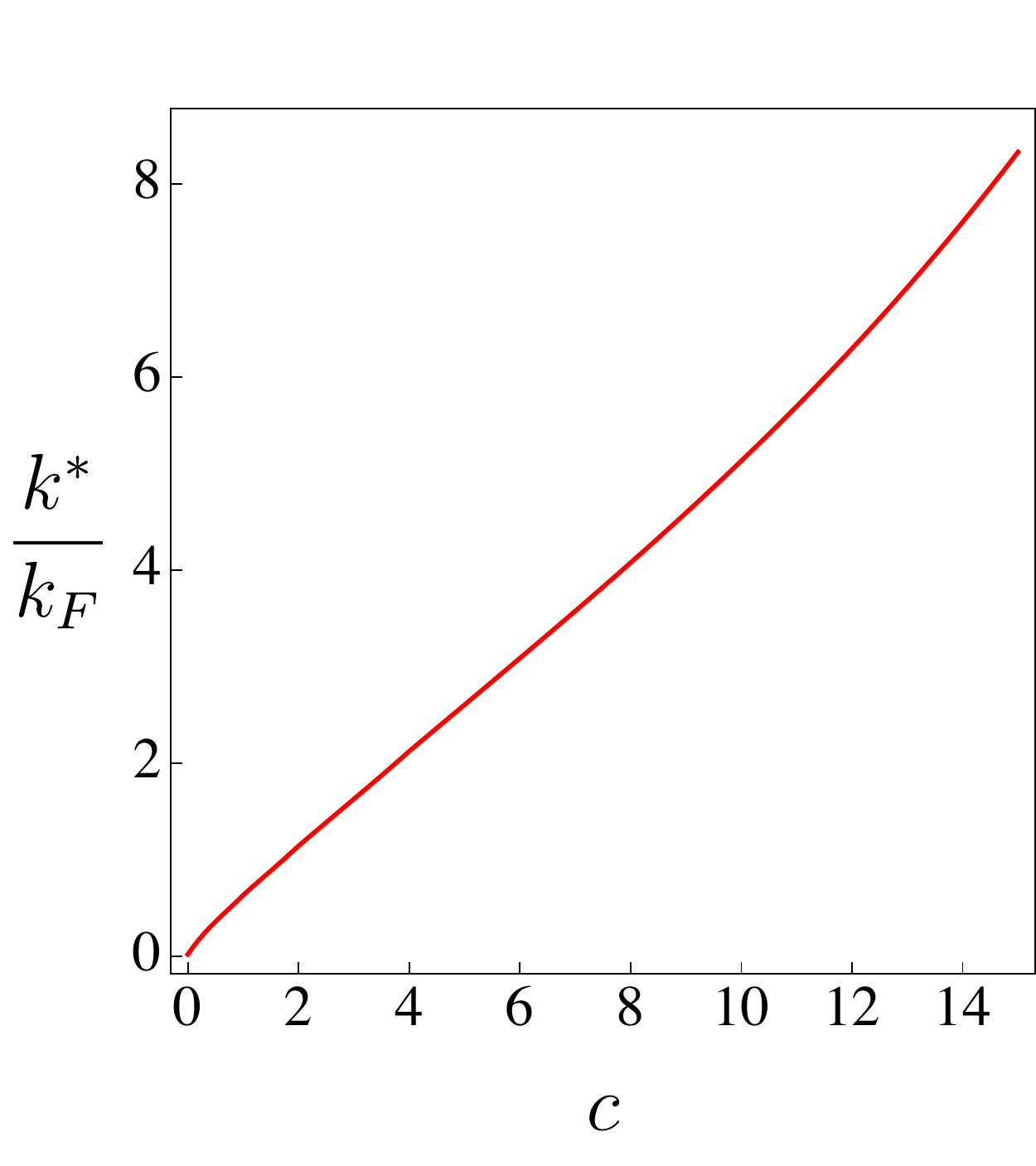}
  \caption{Plot of the dressing threshold~\eqref{bound} on the momentum below which DSF of the ground state has a series expansion in number of particle-hole excitations. The red dots are obtained by numerically computing the bound and the blue line is a guide.}
  \label{fig:bound}
\end{figure}

The situation changes if the backflow function $F_0(q)$ is small. The singularity appearing in $B_m$ is integrable if $1 > 2 F_0(q) > -1$. We can turn this condition on the backflow for a condition on the momentum. For small $k$ the backflow simplfies
\ea {
  F_0(\lambda) = - (p_0 - h_0)\frac{ \hat{L}_{\vartheta}(h_0, \lambda)}{\vartheta(\lambda)},
}
and the momentum is directly proportional to the particle-hole separation
\ea {
  k = 2\pi (p_0-h_0) \rho_t(h_0).
}
Placing the hole directly at the edge, $h_0 = q$, leads to the following condition
\ea {
  |k| < k^{*} =  \Big| \frac{2\pi \rho_t(q)}{\hat{L}_{\vartheta}(q,q)}  \Big| . \label{bound}
} 
The right hand side depends only on the interaction parameter $c$. For large $c$ the dressing threshold is linear in $c$. While $c$ decreases, so does the bound, around $c=2$ the bound equals $k=k_F$, and vanishes when $c$ aproaches 0, see fig~\ref{fig:bound}. 
When $k$ is less than the dressing threshold~\eqref{bound} the correlation function, even for critical state, is organized in a series~\eqref{S_expansion} with number of particle-hole pairs setting the leading power of momentum $k$ of each contribution (as shown in section~\ref{two_ph_and_more}). Above the threshold the relation between number of particle-hole pairs and momentum breaks down and the thermodynamic form factors needs to be dressed with soft modes excitations.

The result of this section shows the existence of a region of small momenta in which the form factors~\eqref{FF1} can be used to compute correlation function even for critical state. Note that this does not mean that in this region a single particle-hole excitation is enough to saturate the correlation function. It only means that there is no dressing of the form factors by soft modes around the discontinuities of the filling function. This results confirms the validity of computation of small momentum limit of the ground state static structure factor presented in~\cite{SciPostPhys.1.2.015}. Motivated by this result, in the next section we derive the edge exponents of the ground state dynamic structure factor in the small momentum limit.

\section{Dynamical correlations of the ground state in the small momentum limit} \label{sec:edge}

In this section we consider the ground state dynamic structure factor. The ground state DSF has a characteristic behaviour along the Lieb I and II modes \cite{RevModPhys.84.1253,PhysRevLett.100.206805}. The Lieb I and II modes correspond to creating a particle-hole excitation with either hole (Lieb I mode) or particle (Lieb II mode) right at the edge of the Fermi sea. As we have seen in the previous section, form factors with excitations placed in the vicinity of the discontinuity of the filling function have a singular behaviour. The singular behaviour of the form factors leads to the singular behaviour of the correlation function, the so-called edge singularities.

As argued in the previous section our approach is not suited for critical states unless we focus on the small momentum DSF with $k$ less than the dressing threshold~\eqref{bound}. Here we go one step further and consider only a single particle-hole excitation. We will show that the ground state correlation function has in this limit the same behavior in the vicinity of edges as predicted by the non-linear Luttinger liquid.

The theory of non-linear Luttinger liquids predicts the following structure of the DSF in the vicinity of the edges.  We denote $\epsilon_{1,2}(k)$ the dispersion relation of the Lieb I and II modes.  
 In the vicinity of the Lieb II mode, $\delta\omega = \omega -\epsilon_2(k) \sim 0$, the DSF has a one sided singularity
\ea {
  S_{\hat{\rho}}(k,\omega) = \theta(\delta\omega) \,\frac{2\pi S_2(k) \left(\delta\omega\right)^{\tilde{\mu}_R + \tilde{\mu}_L - 1}}{\Gamma(\tilde{\mu}_R + \tilde{\mu}_L)(v + v_s)^{\tilde{\mu_L}}|v - v_s|^{\tilde{\mu}_R}},
}
whereas in the vicinity of the Lieb I mode, $\delta\omega = \omega -\epsilon_1(k) \sim 0$, it has a two sided singularity
\ea {
  S_{\hat{\rho}}(k,\omega) = \frac{\theta(\delta\omega)\sin \pi \tilde{\mu}_L + \theta(-\delta\omega)\sin \pi \tilde{\mu}_R }{\sin \pi(\tilde{\mu}_R + \tilde{\mu}_L)}\frac{2\pi S_1(k) \left(\delta\omega\right)^{\tilde{\mu}_R + \tilde{\mu}_L - 1}}{\Gamma(\tilde{\mu}_R + \tilde{\mu}_L)(v + v_s)^{\tilde{\mu_L}}|v - v_s|^{\tilde{\mu}_R}}.
}
In these expressions the exponents $\tilde{\mu}_{R,L}$ are predicted by the non-linear Luttinger liquid \cite{PhysRevLett.100.206805} to~be
\ea {
  \tilde{\mu}_{R,L} = \left(\frac{\sqrt{K}}{2} \mp \frac{1}{2\sqrt{K}} + F(\pm q|\lambda) \right)^2,
}
and $\lambda$ corresponds to the position of particle excitation for Lieb I mode and of the hole excitation for the Lieb II mode.
The velocity $v_s = \partial \epsilon_{1,2}(k)/\partial k$ is the group velocity along the Lieb and it is the same for the Lieb I and II modes.
Parameter $K$ is the Luttinger liquid parameter which for the Lieb-Liniger model is related to the density of the particles
\ea { \label{Luttinger}
  2\pi \rho(q) = \sqrt{K}.
}
Functions $S_{1,2}(k)$ are non-universal prefactors and are related to the thermodynamic limit of ground state form factors. For the Lieb-Liniger model the relation is the following \cite{2012_Shashi_PRB_85_1} 
\ea {
  S_{1,2}(k) = \lim_{N,L\rightarrow\infty} L \left(\frac{L}{2\pi} \right)^{\tilde{\mu}_R + \tilde{\mu}_L} |\langle {\rm GS} |\hat{\rho}(0)|{\rm GS} +  h \to p\rangle|^2,
}
where the particle-hole pair $h \to p$ corresponds to the Lieb I or Lieb II excitation. Their exact expressions where computed in~\cite{2012_Shashi_PRB_85_1}. In the small momentum limit the exponents become
\ea {
  \tilde{\mu}_R &= 1 \mp 2 \frac{k}{k'(q)}\hat{L}_{\vartheta}(q,q) + \mathcal{O}(k^2),\\
  \tilde{\mu}_L &= \mathcal{O}(k^2),
}
where $-$ ($+$) sign corresponds to the exponents in the vicinity of the Lieb I (II) mode.
Notice that since $\hat{L}_{\vartheta}(q,q) = - \partial_q F(q,q)$ the exponents $\tilde{\mu}_R$  can be related to the dressing threshold \eqref{dressing_th} $k^{*}= -\pi \rho_t(q)/\partial_q F(q,q) = \pi \rho_t(q)/\hat{L}_{\vartheta}(q,q)$ 
\ea {
  \tilde{\mu}_R &= 1 \mp k/k^{*} + \mathcal{O}(k^2).
}
The Lieb I and II excitations have the following dispersion relations
\ea {
  \epsilon_{1,2}(k) = kv\left( q \pm \frac{1}{2}\frac{k}{k'(q)}\right) + \mathcal{O}(k^3).
}\label{disprelationsmallk}
The prefactors $S_{1,2}(k)$ in the small momentum are the same and equal \cite{2012_Shashi_PRB_85_1}
\ea {
  S_{1,2}(k) = \frac{K}{2\pi} + \mathcal{O}(k). 
}
This leads to the following form of the DSF at small momentum in the vicinity of the Lieb I and II modes
\ea {
  S_{\rho}(k,\omega) &= \theta(\mp \delta\omega) \,\frac{K}{|v - v_s|^{1\mp 2kL(q,q)/k'(q)}} \left(\delta\omega\right)^{\mp 2kL(q,q)/k'(q)} \nn \\&=  \theta(\mp \delta\omega) \,\frac{K}{|v_s' k/k'(q)|^{1\mp k/k^{*}}} \left(\delta\omega\right)^{\mp k/k^{*}}, \label{finalprefactornLL}
}
where we used that $v(q + k/k'(q)) - v(q) = v'(q)  k/k(q) = v'_s  k/k(q)$. 
We will now show that we can obtain the same results using our approach.

 
We consider the DSF at small momentum, given by a single particle-hole excitation~\eqref{single_particle_DSF}
\begin{equation}
S^{\text{1ph}}_{\hat{\rho}}(k,\omega) = (2\pi)^2\frac{ \rho(h) \rho_h(h + \frac{ k}{k'(h)} ) }{ |k v'(h)  k'(h)|}     |\langle \vartheta| \hat{\rho}(0)|\vartheta, h \to p \rangle |^2 \Big|_{ k v(h + k/k'(h)) = \omega} \;.
\end{equation}
At $T=0$ this formula gives a non-zero result in the range $\epsilon_{2}(k)\leq \omega \leq \epsilon_{1}(k)$. 
The single particle-hole form factors is shown in formula~\eqref{singlep_h_FF}. 
The filling function is discontinuous and therefore the principal value integral in $\mathcal{B}(\vartheta,[p,h])$ has a singular behavior. In the small momentum limit, $p\rightarrow h$, and
\ea {
  \exp\left(\mathcal{B}(\vartheta,[p,h])\right) = \left(\frac{q-h}{q-p}\right)^{-k \hat{L}_{\vartheta}(q,q)/k'(q)}.
}
The limit of the rest of the form factors is simple giving
\begin{equation}
 \lim_{p \to h} |\langle \vartheta| \hat{\rho}(0) |\vartheta, h \to p \rangle | = k'(q) \left(\frac{q-h}{q-p}\right)^{k \hat{L}_{\vartheta}(q,q)/k'(q)}.
\end{equation}
Therefore the correlation function, in the leading order in $k$, is
\ea {
  S^{\text{1ph}}_{\hat{\rho}}(k,\omega) = (2\pi)^2\frac{ \rho(h) \rho_h(h + \frac{ k}{k'(h)} ) }{ |k v'(q)/k'(q)|} \left(\frac{q-h}{q-p}\right)^{-2k \hat{L}_{\vartheta}(q,q)/k'(q)}.
}
The difference between positions of particle and hole and the Fermi momentum $q$ can be expressed in terms of the distance of the energy $\omega$ from the Lieb modes using \eqref{disprelationsmallk}. We obtain
\ea {
 | \omega - \epsilon_2(k)| = | (p-q) k v_s'  | , \qquad |\omega - \epsilon_1(k)| = |(h-q) k v_s'  |.
}
Using that $2 \pi \rho(q) = \sqrt{K} $ and $\lim_{\epsilon \to 0^+}2 \pi  \rho_h(q+\epsilon) = \sqrt{K}$ we arrive to the result
\ea {
  S_{\hat{\rho}}(k,\omega) = \theta(\mp \delta \omega)  \ \frac{ K   }{k v_s' /k'(q)}  \Big| \frac{\delta \omega }{ k v_s'/k'(q)}\Big|^{\mp k/k^{*}}.
  }
which coincides with the non-linear Luttinger liquid result \eqref{finalprefactornLL}.

\section{Numerical evaluations}\label{sec:ABACUS}
In figure \ref{fig:DSF} we compare  the numerical data provided by the \texttt{ABACUS} algorithm \cite{PhysRevA.89.033605} with our analytic result for DSF obtained by the exact single particle-hole contribution \eqref{single_particle_DSF}. Notice that some extra steps are necessary in order to evaluate the Fredholm determinant in formula \eqref{singlep_h_FF}, see Appendix~\ref{sec:freddet}. We have set $c=4$, $m=1$ and two different values of inverse temperature, $\beta=1$ and $\beta=2$ (in units of Fermi energy $\omega_F = k_F^2= (\pi n)^2$), see Fig. \ref{fig:DSF}. \texttt{ABACUS} data are obtained using a finite system size, namely $L=50$ for $\beta=1$ and $L=80$ for $\beta=2$.  Both \texttt{ABACUS} data and our results are convoluted in energy with a Normal distribution of variance $1$ and zero mean.  In order to check for convergence we compute the f-sum rule saturation
\begin{equation}\label{fsumrulesatur}
100 \left( \frac{\int_{0}^{+\infty}  \frac{d\omega}{2 \pi } S_{\hat{\rho}}(k,\omega) \omega (1-e^{-\beta \omega}) }{k^2} \right) \%,
\end{equation}
both for the single particle-hole contribution $S^{\text{1ph}}_{\hat{\rho}}(k,\omega)$ and the data obtained from the  \texttt{ABACUS} algorithm, see Table \ref{tab:fsumrule}. The single particle-hole contribution  to the DSF is shown to well reproduce the full DSF $S_{\hat{\rho}} (k,\omega)\simeq S^{\text{1ph}}_{\hat{\rho}}(k,\omega)$ at small values of $k/k_F$. 
Quite remarkably the figure shows that $S^{\text{1ph}}_{\hat{\rho}}(k,\omega)$ does not only give $S_{\hat{\rho}}(k,\omega)$ at small momenta but it seems to be able also to fully capture the low energy tail even at high values of momentum $k/k_F \sim 1$.

%
\begin {table}[h]
 \caption { {  f-sum rule saturation \eqref{fsumrulesatur}. If the value is $100\%$ it means that all the excitations contributing to the DSF are taken into account. The value of $ 101 \%$ of the \texttt{ABACUS} signals a small $\mathcal{O}(1/L)$ error in the discretization of the thermal distribution with a finite system size $L$.} } \label{tab:fsumrule} 
  \begin{center}
\begin{tabular}{ |p{3cm}||p{3cm}|p{3cm} | }
 \hline
 f-sum rule &   $S^{\text{1ph}}_{\hat{\rho}}(k,\omega)$ &  $S_{\hat{\rho}}(k,\omega)$ \texttt{ABACUS}  \\
 \hline
 $k=\frac{k_F}{5}$, $\beta=2$   & $ 95 \%$    & $100 \%$ \\
 $k=  k_F$, $\beta=2$&    $ 42 \%$  & $ 99 \%$    \\
  \hline
 $k=\frac{k_F}{5}$, $\beta=1$& $97 \%$ & $101 \%$ \\
 $k= k_F$, $\beta=1$    & $57 \%$ & $99 \%$ \\
 \hline
\end{tabular}
\end{center}
\end{table}

\section{Conclusions}

In this paper we reviewed and extended our two works of the past two years \cite{Smooth_us,SciPostPhys.1.2.015} where we studied the thermodynamic form factors of the density operator. We showed here that these form factors can be defined on any non-critical state specified by a given filling function $\vartheta(\lambda)$ and on the ground state, provided that the momentum of the excitations is smaller than the dressing threshold. Our numerical evaluations show that the single particle-hole contribution provides a good approximation to the full DSF at low momenta. Although some numerical result were shown also in \cite{Smooth_us}, thanks to the recent analytical progress we were able to push the numerical evaluations to a higher degree of precision. { We expect our conclusions to be true for the DSF of any local operator $\hat{q}(x)$ that conserves particle number $[\hat{N},\hat{q}(x)]=0$. }

We stress that the analytic knowledge of the dynamical correlation function has multiple applications. First one can easily include the effect of a inhomogenous potential in the gas (like a confining trap) by means of the local density approximation. As our expression of the form factors are functions of the density $n$ (via the function $\vartheta(\lambda)$) one can introduce a local filling function $\vartheta_{x}(\lambda)$ and compute the dynamical correlations at any $x$, analogously to the cases of inhomogenous Luttinger liquids \cite{SciPostPhys.2.1.002,SciPostPhys.3.3.019,1712.05262}. Moreover with our approach one can compute the DSF in the low momentum after a homogeneous quantum quench.  On the other hand inhomogenous initial states present non-equilibrium steady states with $\vartheta(\lambda)$ a discontinuous function of $\lambda$ \cite{Ben,arXiv.1605.09790,1712.05262}. Therefore we should expect that also for this case the form factors need to be dressed with soft modes excitations with momenta close to the discontinuity.

Many question are still open. The recently introduced generalized hydrodynamics (GHD) \cite{arXiv.1605.09790,Ben,PhysRevB.96.115124,PhysRevLett.119.220604,PhysRevLett.119.195301,SciPostPhys.3.3.020,Doyon2017} has suggested a universal form for the low momentum limit of the form factors of operators  $\hat{q}(x)$ that are globally conserved. Since this form is independent of the model and only depends on the resolvent $\hat{L}_{\vartheta}(\lambda,\mu)$ it is reasonable to ask if the same degree of universality applies to the full particle-hole form factors.  As the density operator belong to this class of operators (as the total density $n= \int \langle \hat{\rho}(x) \rangle dx$ is conserved by the Lieb-Liniger Hamiltonian), one could try to guess the thermodynamic form factors for any operator density $\hat{q}(x)$ from our result and from the other information given by GHD. Moreover other simplifications in our formula for the form factors could be found, such as their expression could be easily extended to different models and operators. This would allow to extend calculations to models with bound states excitations, like the XXZ spin chain whose exact computation of thermodynamic form factors is at the moment still an open problem. 

Furthermore we mention that while the correlations given by a single particle-hole excitation are connected to the presence of a large scale generalized hydrodynamics,  the contribution from higher numbers of particle-hole pairs could teach how to add viscous and dissipative terms to the hydrodynamic description \cite{Boldrighini1997,1742-5468-2017-7-073210,SciPostPhys.2.2.014,1742-5468-2016-6-064005,SciPostPhys.3.5.033,PhysRevB.96.220302}. Therefore a more extensive analysis of the two particle-hole contribution is necessary. 

The dressing of the form factors when the reference state is critical is still an open question for our approach. It seems that in order to dress the form factors with its soft modes one has to choose a proper regularization and sum over them. One way to proceed is to go back to the finite size $L$, introduce the quantum numbers of the soft modes and sum analytically over them before to take the thermodynamic limit. This is indeed what was done in \cite{1742-5468-2012-09-P09001}. Clearly this procedure is very complicated due to the sum over quantum numbers, and it would be desirable to be able to start directly from the thermodynamic form factors, introduce some other type of regularization and sum over the soft modes.  

{ Finally it would be interesting to understand if a single particle-hole excitation saturates also the dynamical correlations in generic, non-integrable, gapless models. This could be tested for example with a variational density matrix approach \cite{PhysRevLett.111.020402,PhysRevB.92.125136}. }

\section*{Acknowledgements}
 JDN and MP kindly acknowledge Jean-S\'{e}bastien Caux for the use of the \texttt{ABACUS} algorithm and for extremely valuable comments on the manuscript.



\paragraph{Funding information}
The authors acknowledge support from LabEx ENS-ICFP:ANR-10-LABX-0010/ANR-10-IDEX-0001-02 PSL* (JDN) and from the NCN under FUGA grant \mbox{2015/16/S/ST2/00448} (MP).

\section*{References}

\bibliographystyle{iopart-num}

\bibliography{thermo_ff}

\begin{appendices}

\section{Role of the averaging state}\label{app:averaging_state}

In the original paper~\cite{Smooth_us} we have assigned to a given smooth distribution $\vartheta$ a microscopic state, which we called the averaging state. The form factors in the thermodynamic limit should not depend on this microscopic choice of the averaging state. Under this assumption we could choose the averaging state to suit the best our purpose: computation of the thermodynamic limit of the form factors. In this appendix we discuss how the specific choice of the averging form factors influences the computations.

In the expression for the finite-size form factors there are terms that require certain care while taking the thermodynamic limit. These are expression of the form
\begin{equation}
  \frac{1}{\lambda_j - \lambda_k},
\end{equation}
with the two rapidities close to each other. In the finite system the rapidities are always distinct and these kind of terms are finite. In the thermodynamic limit, the distance between the neigbouring rapidities behaves like $1/L$ and therefore this terms potentially diverge for large $L$. 

Consider the difference between two neigbouring rapidities. Using the Bethe equations~\eqref{bethe} we have
\begin{align}
  \frac{2\pi}{L}\left( I_j - I_k\right) &= \lambda_j - \lambda_k + \int_{-\infty}^{\infty} {\rm d}\lambda \left(\theta(\lambda_j - \lambda) - \theta(\lambda_k - \lambda)\right) \rho(\lambda) + \mathcal{O}(1/L) \nonumber \\
  &= \left(\lambda_j - \lambda_k\right)\left( 1 + \int_{-\infty}^{\infty} {\rm d}\lambda K(\lambda_j - \lambda) \rho(\lambda) \right) + \mathcal{O}((\lambda_j-\lambda_k)^2).
\end{align}
The expression in the bracket is $2\pi \rho_t(\lambda_j)$, see eq.~\eqref{rho_total}, and therefore
\begin{equation}
  \lambda_j - \lambda_k = \frac{I_j - I_k}{L \rho_t(\lambda_j)}.
\end{equation}
The quantum numbers, or a difference between them, depends on the microscopic configuration of the averaging state. One convenient realization is defined in the following way.

The uniform averaging state is the one in which for each interval $[\lambda, \lambda + {\rm d}\lambda]$ the rapidities are distributed uniformly. If the rapidities are distributed uniformly the corresponding numbers are also distributed uniformly and the difference between them is proportional to the difference in indices. The proportionality constant is the inverse of the filling function $\vartheta(\lambda)$. For example, if the filling function is $1/2$ than every second quantum number must be occupied, so the distance between two consecutive quantum numbers is $2$. Therefore
\begin{equation}
  \lambda_j - \lambda_k = \frac{j-k}{L \rho(\lambda_j)}.
\end{equation}
The thermodynamic limit of the form factors was derived as a thermodynamic limit for such uniform averaging state. This has some consequences on the structure of the resulting form factors.

In the expressions for the Fredholm determinants and integral equations for function $W$ there are divergences of the form $1/(h-\lambda)$. These divergences comes from computing thermodynamic limit of the finite sums of the following form
\begin{equation}
  \frac{1}{L}\sum_{j \neq k} \frac{f(\lambda_j)}{(\lambda_k - \lambda_j)}.
\end{equation}
To take the thermodynamic limit we split the sum in two regions, one where the difference $|\lambda_k - \lambda_j|$ is large and the second where this difference is small. We introduce
\begin{equation}
  \nu^*(\lambda_k) = \frac{n^*}{L\rho_p(\lambda_k)},
\end{equation}
as a cutoff between these two regions. In the first region we can safely take the thermodynamic limit to get integrals
\begin{equation}
  \frac{1}{L}\sum_{j \neq [k-n^*, k+n^*]} \frac{f(\lambda_j)}{(\lambda^- - \lambda_j)} = \int_{-\infty}^{\lambda_k - \nu(\lambda_k)} {\rm d}\lambda \rho(\lambda)\frac{f(\lambda)}{\lambda_k - \lambda} + \int_{\lambda_k + \nu(\lambda_k)}^{\infty} {\rm d}\lambda \rho(\lambda)\frac{f(\lambda)}{\lambda_k - \lambda}.
\end{equation}
In the second region we substitute the difference of rapidities with a difference of quantum numbers.
\begin{equation}
  \frac{1}{L}\sum_{\substack{j = k-n^*\\ j \neq k }}^{k+n^*} \frac{f(\lambda_j)}{\lambda_k - \lambda_j} = \rho_t(\lambda_k) \sum_{\substack{j = k-n^*\\ j \neq k }}^{k+n^*}  \frac{f(\lambda_j)}{I_k - I_j}.
\end{equation}
Under the assumption that $f(\lambda)$ is a smooth function, in the leading order in $\nu^*(\lambda_k)$ we get 
\begin{equation}
  \frac{1}{L}\sum_{\substack{j = k-n^*\\ j \neq k }}^{k+n^*} \frac{f(\lambda_j)}{\lambda_k - \lambda_j} = \rho_t(\lambda_k)f(\lambda_k) \sum_{\substack{j = k-n^*\\ j \neq k }}^{k+n^*}  \frac{1}{I_k - I_j} = \rho_t(\lambda_k)f(\lambda_k) \sum_{j=1}^{n^*} \left( \frac{1}{I_k - I_{k-j}} + \frac{1}{I_k - I_{k+j}}\right).
\end{equation}
The evaluation of the sum depends on the details of the averaging state. However, if for the averaging state we choose the uniform state, than the sum is zero because for the uniform state
\begin{equation}
  I_k - I_{j} = \vartheta(\lambda_k) (k-j) + \mathcal{O}((\lambda_k - \lambda_j)^2).
\end{equation}
Therefore the sum is equal to the sum of the two integrals and in the thermodynamic limit $\nu^*(\lambda_k) \rightarrow 0$ leads to the principal value integral defined in~\eqref{Pint} 
\ea {
  \frac{1}{L}\sum_{j \neq k} \frac{f(\lambda_j)}{(\lambda_k - \lambda_j)} \rightarrow \pint_{-\infty}^{\infty} {\rm d}\lambda \frac{f(\lambda)}{\lambda_k - \lambda}.
}

There are is also another part of the form factors which depends crucially on the microscopic of the averaging state. This is the double product
\begin{equation}
  \prod_{j\neq k} \left(\frac{(\lambda_j - \lambda_k)(\mu_j-\mu_k)}{(\mu_k - \lambda_j)} \right)^{1/2}.
\end{equation}
In this product whenever the rapidities are close to each other we can substitute for their difference the difference of quantum numbers. For example, the computations lead us to consider the following product
\begin{equation}
  \prod_{k=j-n^*}^{j-1} (I_j - I_k).
\end{equation}
Again, if the quantum numbers are distributed uniformly than the product is proportional to a factorial
\begin{equation}
  \prod_{k=j-n^*}^{j-1} (I_j - I_k) = \vartheta(\lambda_k)^{-n^*+1} \frac{\Gamma(n^*+1)}{\Gamma(1)}.
\end{equation}
If the quantum numbers are not distributed uniformly, than there are elements of the product that deviate from the factorial structure and if we insist in expressing the answer through factorial they lead to extra multiplicative terms. These extra terms are $1$ if the state is uniform and different from $1$ otherwise. In the form factors there are more expression like this and one would need to carefully collect all of them. Under the assumption that the thermodynamic limit of the form factors does not depend on the microscopic structure of the averaging state all these extra terms need to cancel each other. At present we do not attempt to prove this statement.

\section{Derivations}\label{app:derivation}

In this appendix we provide derivations of some formulas from the main text. The first three subsections concern with a derivation of the new formula for the form factors presented in section~\ref{sec:new_expression}. In the last subsection we derive an expression for the form factors when the excited state consists of a single dominant excitation and a number of small excitations (soft modes). This result is used in section~\ref{sec:dressing}.

\subsection{Fredholm determinant}

We start by rewriting the Fredholm determinant ${\rm Det}(\boldsymbol{1} -\hat{A})$ from the numerator of~\eqref{FF_D} using the generalized resolvent $L^{[\mathbf{p}, \mathbf{h}]}$ defined in eq.~\ref{L_ph_def}.
We multiply the Fredholm determinant ${\rm Det}(\boldsymbol{1} -\hat{A})$ by $1$ expressed as
\ea {
  {\rm Det}(\boldsymbol{1} -K^{[\mathbf{p}, \mathbf{h}]}){\rm Det}(\boldsymbol{1} + L^{[\mathbf{p}, \mathbf{h}]}) = 1.
}
The result is
\ea {
  {\rm Det}\left( \boldsymbol{1} - \tilde{a}^{[\mathbf{p}, \mathbf{h}]}\left(K- \frac{2}{c}\right)\right) = {\rm Det}\left( \boldsymbol{1}-K^{[\mathbf{p}, \mathbf{h}]}\right) {\rm Det}\left(\boldsymbol{1} + \mathcal{L}_a\right),
}
with
\ea {
  \mathcal{L}_a(\lambda,\lambda') = \left(\boldsymbol{1} + \int_{-\infty}^{\infty} {\rm d}\alpha L^{[\mathbf{p}, \mathbf{h}]}(\lambda,\alpha) \right) \tilde{a}^{[\mathbf{p}, \mathbf{h}]}(\lambda')\frac{2}{c} = \frac{1}{\pi c} \rho_t^{[\mathbf{p},\mathbf{h}]}(\lambda)\tilde{a}^{[\mathbf{p}, \mathbf{h}]}(\lambda').
}
Function $\rho_t^{[\mathbf{p},\mathbf{h}]}(\lambda)$ was defined in eq.~\eqref{rho_total_ph}.
The kernel $\mathcal{L}_a$ is seperable and thus
\ea {
  {\rm Det}\left(\boldsymbol{1} + \mathcal{L}_a\right) = 1 + \frac{1}{\pi c}\int_{-\infty}^{\infty} {\rm d}\lambda \rho_t^{[\mathbf{p},\mathbf{h}]}(\lambda) \tilde{a}^{[\mathbf{p}, \mathbf{h}]}(\lambda) = 1 + \frac{2 n^{[\mathbf{p}, \mathbf{h}]}}{c}, 
}
where we use definition of $n^{[\mathbf{p}, \mathbf{h}]}$ from eq.~\eqref{n_ph}.
The answer for the Fredholm determinant~is
\ea {
  {\rm Det}\left( \boldsymbol{1}- \tilde{a}^{[\mathbf{p}, \mathbf{h}]}\left(K- \frac{2}{c}\right)\right) = \left(1 + \frac{2n^{[\mathbf{p}, \mathbf{h}]}}{c} \right) {\rm Det}\left( \boldsymbol{1} -K^{[\mathbf{p}, \mathbf{h}]}\right).
}

\subsection{Function \texorpdfstring{$W(h_i,  \lambda)$}{W(h, lambda)}}

Function $W_i(\lambda) \equiv W(h_i, \lambda)$ obeys an integral equation~\eqref{eq_W} which we repeat here
\ea {
  W_i(\lambda) - \pint_{-\infty}^{\infty} {\rm d}\alpha W_i(\alpha) \tilde{a}^{[\mathbf{p}, \mathbf{h}]}(\alpha)\left(K(\alpha - \lambda) - \frac{2}{c}\right) = b_i\left(K(h_i-\lambda) - \frac{2}{c} \right),
}
with
\ea {
  b_i = - \frac{\resatilde^{[\mathbf{p}, \mathbf{h}]}}{\vartheta(h_i)F(h_i)}.
}
To solve this equation we first split it in two simpler equations for two new functions defined through
\ea { \label{W_split}
  W_i(h,\lambda) = - \frac{2b_i}{c} \left(W_{i,1}(\lambda) + W_{i,2}(\lambda) \right),
}
where
\ea {
  W_{i,1}(\lambda) - \pint_{-\infty}^{\infty} {\rm d}\alpha W_{i,1}(\alpha) \tilde{a}^{[\mathbf{p}, \mathbf{h}]}(\alpha) \left(K(\alpha - \lambda) - \frac{2}{c} \right) &= 1,\\
  W_{i,2}(\lambda) - \pint_{-\infty}^{\infty} {\rm d}\alpha W_{i,2}(\alpha) \tilde{a}^{[\mathbf{p}, \mathbf{h}]}(\alpha) \left(K(\alpha - \lambda) - \frac{2}{c} \right) &= -\frac{c}{2}K(h_i-\lambda). \label{eq_W_2}
}
We solve for the first function first. We make an ansatz
\ea {
  W_{i,1}(h,\lambda) = d_1 \rho_t^{[\mathbf{p},\mathbf{h}]}(\lambda), 
}
to find
\ea {
  d_1 = \frac{2\pi}{1 + \frac{2}{c}n^{[\mathbf{p},\mathbf{h}]}}.
}
For $W_{i,2}(\lambda)$ we solve in two steps. First we write
\ea { \label{W_2}
  W_{i,2}(\lambda) = \bar{W}_{i,2}(\lambda) + d_2 \frac{L^{[\mathbf{p}, \mathbf{h}]}(h_i,\lambda)}{2\pi \tilde{a}^{[\mathbf{p}, \mathbf{h}]}(\lambda)} ,
}
and use that the combination ${L^{[\mathbf{p}, \mathbf{h}]}(\lambda,\lambda')}/(2\pi \tilde{a}^{[\mathbf{p}, \mathbf{h}]}(\lambda')) $ solves the following integral equation
\ea {
  \frac{L^{[\mathbf{p}, \mathbf{h}]}(\lambda,\lambda')}{2\pi \tilde{a}^{[\mathbf{p}, \mathbf{h}]}(\lambda')}  - \pint_{-\infty}^{\infty}{\rm d}\alpha\tilde{a}^{[\mathbf{p}, \mathbf{h}]}(\alpha) \left(\frac{L^{[\mathbf{p}, \mathbf{h}]}(\lambda, \alpha)}{2\pi \tilde{a}^{[\mathbf{p}, \mathbf{h}]}(\alpha)}  \right) K(\alpha, \lambda') = \frac{1}{2\pi} K(\lambda - \lambda').
}
Inserting expression~\eqref{W_2} into  integral equation~\eqref{eq_W_2} we find
\ea {
  \bar{W}_{i,2}(\lambda) - \pint_{-\infty}^{\infty} {\rm d}\alpha \bar{W}_{i,2}(\alpha) \tilde{a}^{[\mathbf{p}, \mathbf{h}]}(\alpha) \left(K(\alpha - \lambda) - \frac{2}{c} \right) = -\left(\frac{d_2}{2\pi} + \frac{c}{2} \right)K(h_i-\lambda) \nonumber\\
  - \frac{d_2}{2\pi} \frac{2}{c}\pint_{-\infty}^{\infty} {\rm d}\alpha L^{[\mathbf{p}, \mathbf{h}]}(h_i,\alpha).
}
Fixing $d_2 = -\pi c$ the right hand side becomes $\lambda$ independent. Moreover the remaining integral can be expressed through $\rho_t^{[\mathbf{p},\mathbf{h}]}(\lambda)$ defined in~\eqref{rho_total_ph}. The result is  
\ea {
  \bar{W}_{i,2}(\lambda) - \pint_{-\infty}^{\infty} {\rm d}\alpha \bar{W}_{i,2}(\alpha) \tilde{a}^{[\mathbf{p}, \mathbf{h}]}(\alpha) \left(K(\alpha - \lambda) - \frac{2}{c} \right) = 2\pi \rho_t^{[\mathbf{p},\mathbf{h}]}(h_i) - 1. 
}
This has a solution of the form
\ea {
  \bar{W}_{i,2}(\lambda) = d_3(h_i) \rho_t^{[\mathbf{p},\mathbf{h}]}(\lambda),
}
with
\ea {
  d_3(h_i) = \frac{2\pi}{1 + \frac{2}{c}n^{[\mathbf{p},\mathbf{h}]}}\left(2\pi\rho_t^{[\mathbf{p},\mathbf{h}]}(h_i) - 1 \right).
}
For the sum of $W_{i,1}(\lambda)$ and $W_{i,2}(\lambda)$ we find
\ea {
  W_{i,1}(\lambda) + W_{i,2}(\lambda) = -\frac{c}{2}\frac{L^{[\mathbf{p}, \mathbf{h}]}(h_i,\lambda)}{\tilde{a}^{[\mathbf{p}, \mathbf{h}]}(\lambda)} + \frac{2\pi \rho_t^{[\mathbf{p},\mathbf{h}]}(\lambda)\, 2\pi \rho_t^{[\mathbf{p},\mathbf{h}]}(h_i)}{1 + \frac{2}{c}n^{[\mathbf{p},\mathbf{h}]}},
}
and using eq.~\eqref{W_split} 
\ea {
  W(h_i, \lambda) = \frac{\resatilde^{[\mathbf{p}, \mathbf{h}]}}{\vartheta(h_i)F(h_i)}\left( - \frac{L^{[\mathbf{p}, \mathbf{h}]}(h_i,\lambda)}{\tilde{a}^{[\mathbf{p}, \mathbf{h}]}(\lambda)} + \frac{2}{c} \frac{2\pi \rho_t^{[\mathbf{p},\mathbf{h}]}(h_i) 2\pi \rho_t^{[\mathbf{p},\mathbf{h}]}(\lambda)}{1 + \frac{2}{c}n^{[\mathbf{p},\mathbf{h}]}}  \right).
}
Both $L^{[\mathbf{p}, \mathbf{h}]}(h_i,\lambda)$ and $\tilde{a}^{[\mathbf{p}, \mathbf{h}]}(\lambda)$ have a simple pole at $\lambda = h_j$ and therefore their ratio is finite. Therefore
\ea {
  W(h_i,h_j) =  \frac{ \resatilde^{[\mathbf{p}, \mathbf{h}]} }{\vartheta(h_i)F(h_i)} \left( - \left[ \lim_{\lambda \to h_j} \frac{L^{[\mathbf{p}, \mathbf{h}]}(h_i,\lambda)}{\tilde{a}^{[\mathbf{p}, \mathbf{h}]}(\lambda)} \right] +  \frac{2}{c} \frac{2\pi \rho_t^{[\mathbf{p}, \mathbf{h}]}(h_i) 2\pi \rho_t^{[\mathbf{p}, \mathbf{h}]}(h_j)}{1 + \frac{2}{c}n^{[\mathbf{p},\mathbf{h}]}} \right).
}

\subsection{Dependence on constants in the prefactor}\label{sec:constant_alpha}

Let us temporarily introduce the following notation
\ea {
  \alpha &= \frac{2}{c},\\
  \beta &= \frac{\alpha}{1 + \alpha \,n^{[\mathbf{p},\mathbf{h}]}}. 
}
Combining the results of two previous subsections we obtain
\ea { \label{alpha_independence}
  {\rm Det}(\boldsymbol{1}  - \hat{A})\, {\rm det}_{i,j=1}^m \left(\delta_{ij} + W(h_i,h_j) \right) = \left( \boldsymbol{1} -K^{[\mathbf{p}, \mathbf{h}]}\right) \beta^{-1} \det\left(A_{ij} + \beta B_{ij} \right),
}
where we defined the following matrix elements
\ea {
  A_{ij} &=  \delta_{ij}  - \frac{\resatilde^{[\mathbf{p}, \mathbf{h}]}}{\vartheta(h_i)F(h_i)}\left[ \lim_{\lambda \to h_j} \frac{L^{[\mathbf{p}, \mathbf{h}]}(h_i,\lambda)}{\tilde{a}^{[\mathbf{p}, \mathbf{h}]}(\lambda)} \right],\\
  B_{ij} &= \frac{\resatilde^{[\mathbf{p}, \mathbf{h}]}}{\vartheta(h_i) F(h_i)} 2\pi \rho_t^{[\mathbf{p}, \mathbf{h}]}(h_i) 2\rho_t^{[\mathbf{p}, \mathbf{h}]}(h_j).
}
From the finite-size form factors~\cite{2012_Shashi_PRB_85_1} we know that the factor~\eqref{alpha_independence} is independent of $\alpha$ and therefore also of $\beta$. Therefore choosing $\beta=1$ we obtain
\ea {
  {\rm Det}(\boldsymbol{1} - \hat{A})\, {\rm det}_{i,j=1}^m \left(\delta_{ij} + W(h_i,h_j) \right) = \left( \boldsymbol{1} -K^{[\mathbf{p}, \mathbf{h}]}\right) \det\left(A_{ij} + B_{ij} \right),
}
Inserting expressions for $A_{ij}$ and $B_{ij}$ we get the final result reported in eq.~\eqref{eq:newD}.

It is interesting to try to prove that expression~\eqref{alpha_independence} is independent of $\beta$ at the thermodynamic level. For~\eqref{alpha_independence} to be independent of $\beta$, $A_{ij}$ must be of rank $n-1$ and $B_{ij}$ must be of rank $1$ because then
\ea {
  {\rm det}(A_{ij} + \beta B_{ij}) = \beta {\rm det}(A_{ij} + B_{ij}).
}
Matrix $B_{ij}$ has a product form and thus its rank is $1$. It is more difficult to show that the matrix $A_{ij}$ has rank $n-1$. 
Consider the $n=1$ case. Then we must have
\ea {
  \frac{\tilde{a}_{\text{res}(h)}^{[p,h]}}{\vartheta(h)F(h)} \lim_{\lambda\rightarrow h}\frac{L^{[p,h]}(h,\lambda)}{\tilde{a}^{[p,h]}(\lambda)} = 1.
}
We have seen in~\cite{SciPostPhys.1.2.015} that this equality holds in the small momentum limit.
Here, we will show that it is also true  for large $c$. We will keep the first two leading orders in $1/c$. The backflow function is
\ea {
  F(\lambda) = - \frac{p-h}{\pi c} \left(1  + \frac{1}{\pi c}\int_{-\infty}^{\infty} {\rm d}\alpha\, \vartheta(\alpha) + \mathcal{O}(1/c^2) \right).
}
The $\tilde{a}^{[p,h]}(\lambda)$ function is
\ea {
  \tilde{a}^{[p, h]}(\lambda) = \frac{\vartheta(\lambda)}{2\pi} \frac{p-\lambda}{h-\lambda}\left(1 + \frac{p-h}{\pi c} \pint_{\infty}^{\infty}{\rm d}\alpha \frac{\vartheta(\alpha)}{\alpha - \lambda} + \mathcal{O}(1/c^2)\right),
}
and its residue is
\ea {
  \tilde{a}_{\text{res}(h)}^{[p,h]} = - \frac{\vartheta(h)}{2\pi} (p-h)\left(1 + \frac{p-h}{\pi c} \pint_{\infty}^{\infty}{\rm d}\alpha \frac{\vartheta(\alpha)}{\alpha-h} + \mathcal{O}(1/c^2)\right).
}
Therefore the right hand side is
\ea {
  \frac{\vartheta(h)F(h)}{\tilde{a}_{\text{res}(h)}^{[p, h]}} &= \frac{2}{c}\left(1  + \frac{1}{\pi c}\int_{-\infty}^{\infty} {\rm d}\alpha\, \vartheta(\alpha)\right)\left(1 - \frac{p-h}{\pi c} \pint_{\infty}^{\infty}{\rm d}\alpha \frac{\vartheta(\alpha)}{h - \alpha}\right)^{-1} \nonumber \\
  &= \frac{2}{c}\left( 1 + \frac{1}{\pi c}\pint_{-\infty}^{\infty}{\rm d}\alpha\, \vartheta(\alpha)\, \left( 1 + \frac{p-h}{h-\alpha}\right)\right) \nonumber \\
  &= \frac{2}{c}\left( 1 + \frac{1}{\pi c}\pint_{-\infty}^{\infty}{\rm d}\alpha\, \vartheta(\alpha)\,  \frac{p-\alpha}{h-\alpha} + \mathcal{O}(1/c^2)\right).
}
On the other hand the ratio $L^{[p, h]}(\lambda,\lambda')/\tilde{a}^{[p, h]}(\lambda')$ is given by
\ea {
  \frac{L^{[p, h]}(\lambda,\lambda')}{\tilde{a}^{[p,h]}(\lambda')} &= \frac{2}{c} + \left(\frac{2}{c}\right)^2 \pint_{-\infty}^{\infty}{\rm d}\alpha\,  \tilde{a}^{[p, h]}(\alpha) + \mathcal{O}(1/c^3)\nonumber\\
  &= \frac{2}{c}\left(1 + \frac{1}{\pi c} \pint_{-\infty}^{\infty}{\rm d}\alpha\, \vartheta(\alpha)\frac{p-\alpha}{h-\alpha} + \mathcal{O}(1/c^2) \right). 
}
Therefore the identity holds for the first two orders in $1/c$. We showed it also in the small momentum limit of the two-particle form factors, see section \ref{exactcompFF}. Proving that eq.~\eqref{alpha_independence} is independent of $\beta$  in full generality remains an open question.

\subsection{Derivation of formula \texorpdfstring{\eqref{ff_dressing}}{(125)}}\label{derivation_dressing}
Comparing expressions for the form factors with $1$ and $m+1$ particle-hole excitations we find
\ea {
  \frac{\mathcal{A}(\vartheta, \{ h_j,p_j\}_{j=0}^m\})}{\mathcal{A}(\vartheta, \{ h_0,p_0\})} =& \mathcal{A}(\vartheta, \{ h_j,p_j\}_{j=1}^m\})  \prod_{j=1}^m \left[\frac{(p_0 - h_j + ic)^2}{(h_{0,j} + ic)(p_{0,j}+ic)}\frac{(p_j - h_0 + ic)^2}{(h_{j,0} + ic)(p_{j,0}+ic)} \right]^{1/2}\nn\\
  &\times 
  \prod_{j=1}^m \frac{ h_{0,j}\, p_{0,j}}{ (p_j-h_0)(p_0-h_j)}.
}
Let analyze first products involving the dominant excitation. 
For $j=1,\dots, m$ we have
\begin{align}
  \frac{(p_0 - h_j + ic)^2}{(h_0 -h_j +ic)(p_0 - p_j+ic)} &= \frac{p_0 - h_j + ic}{h_0 - h_j + ic}.\\
  \frac{(p_j - h_0 + ic)^2}{(h_j -h_0 +ic)(p_j - p_0+ic)} &= \frac{p_j - h_0 + ic}{p_j - p_0 + ic}. 
\end{align}
The product of these two terms is in the leading order a pure phase factor and its contribution to the form factors can be neglected. To compute the correlation function we need only the norm of the form factors. For $i=0$ and $j$ corresponding to a small excitation we have
\begin{equation}
  \frac{(h_0 - h_j)(p_0-p_j)}{(p_0 - h_j)(p_j - h_0)} = -1 + \mathcal{O}(p_j-h_j).
\end{equation}
Therefore 
\ea {
  \mathcal{A}(\vartheta, \{ h_j,p_j\}_{j=0}^m\}) =& (-1)^m e^{i\phi} \mathcal{A}(\vartheta, \{ h_j,p_j\}_{j=1}^m\}) \times \left(1 + \mathcal{O}(p_j-h_j)\right).
}
We analyze now the term $\mathcal{A}(\vartheta, \{ h_j,p_j\}_{j=1}^m\})$ involving only small excitations.
For $i,j=1,\dots, m$ we have
\ea {
  \frac{(p_i - h_j + ic)^2}{(h_i -h_j +ic)(p_i - p_j+ic)} &= \frac{(h_i - h_j + ic + p_i - h_i)(p_i - p_j + ic + p_j - h_j)}{(h_i -h_j +ic)(p_i - p_j+ic)} \nonumber \\
  &= 1 + \mathcal{O}(p_j-h_j).
}
The double product,
\begin{equation}
  \frac{\prod_{i<j=1}^m (h_i - h_j)(p_i-p_j)}{\prod_{i,j=1}^m(p_i - h_j)},
\end{equation}
remains as it is.
Consider now the ratio of the back-flow functions. The back-flow function is smooth, but the filling function is, for now, arbitrary. We have
\begin{equation}
  \frac{\pi \tilde{F}_0(p_k) \sin \pi \tilde{F}_0(h_k)}{\pi \tilde{F}_0(h_k) \sin \pi \tilde{F}_0(p_k)} = \frac{\vartheta(p_k) \sin \pi \vartheta(h_k) F_0(h_k)}{\vartheta(h_k) \sin \pi \vartheta(p_k)F_0(h_k)}. 
\end{equation}
Finally $\rho_t(\lambda)$ is a smooth function, irrespective of the filling function, and therefore can be evaluated either at $p_i$ or $h_i$. Putting all the pieces together we find
\ea {
  \frac{\mathcal{A}(\vartheta, \{ h_j,p_j\}_{j=0}^m\})}{\mathcal{A}(\vartheta, \{ h_0,p_0\})} = \left[\prod_{k=1}^m \frac{F_0(h_k)}{\rho_t(h_k)} \frac{\vartheta(p_k) \sin \pi \vartheta(h_k) F_0(h_k)}{\vartheta(h_k) \sin \pi \vartheta(p_k)F_0(h_k)} \right] \frac{\prod_{i<j=1}^m (h_i - h_j)(p_i-p_j)}{\prod_{i,j=1}^m(p_i - h_j)}.
}
For the $\mathcal{B}(\vartheta, \{ p_j, h_j\}_{j=0}^m)$ term we find in the leading order $\mathcal{O}(p_j-h_j)$
\ea {
  \mathcal{B}(\vartheta, \{ p_j, h_j\}_{j=0}^m) = \mathcal{B}\left(\vartheta, \{ p_0, h_0\}\right) + \sum_{k=1}^m \pint_{-\infty}^{+\infty} {\rm d}\lambda \frac{\tilde{F}(\lambda)(h_k-p_k)}{(\lambda-h_k)(\lambda-p_k)}.
}
All the other terms in $\mathcal{B}(\vartheta, \{ p_j, h_j\}_{j=0}^m)$ that do not involve summation over excitations depend, in the leading order, only on the backflow $F_0(\lambda)$ of the dominant excitation and therefore are contained in $\mathcal{B}(\vartheta, \{ p_0, h_0\})$. The remaining integral
\ea {
  \int_{-\infty}^{+\infty} {\rm d}\lambda \frac{\tilde{F}(\lambda)(p_k-h_k)}{(\lambda-h_k+ic)(\lambda-p_k+ic)},
}
is proportional to $p_k-h_k$ and therefore does not contribute to the leading order in $p_k-h_k$. It might seem that the same is true for principal value integral. However, as we will see later, the contribution of this integral depends on the smoothness of the filling function $\vartheta(\lambda)$. As we do not require the filling function to be smooth everywhere, we leave this term as it is. 

The last piece is the $\mathcal{D}(\vartheta, \{ p_j, h_j\}_{j=0}^m)$ part. We have
\ea {
  \frac{\mathcal{D}(\vartheta, \{ p_j, h_j\}_{j=0}^m)}{\mathcal{D}(\vartheta, \{ p_0, h_0\})} = \frac{{\rm det}_{i,j=0}^m\left( \delta_{ij} + W(h_i,h_j)\right)}{1 + W_0(h_0,h_0)} \frac{{\rm Det}(\boldsymbol{1}-\hat{A})}{{\rm Det}(\boldsymbol{1}-\hat{A}_0)}.
}
Crucial in the evaluation of this ratio is function $\tilde{a}^{[\mathbf{p}, \mathbf{h}]}(\lambda| \{ p_j,h_j\}_{j=0}^m)$. Inspecting its form we see that in the leading order in $p_k-h_k$ its equal to the $\tilde{a}^{[\mathbf{p}, \mathbf{h}]}(\lambda| \{ p_0,h_0\})$. Therefore the ratio of the Fredholm determinants is equal to $1$.

The remaining part involves matrix elements $W(h_i, h_j)$ given in~\eqref{newW}. The matrix elements are proportional to $\resatilde^{[\mathbf{p},\mathbf{h}]}$ which, for $i=1,\dots,n$ is of order $p_i-h_i$. Therefore the leading order of the determinant is given by the contribution from the dominant excitation and
\ea {
  \frac{\mathcal{D}(\vartheta, \{ p_j, h_j\}_{j=0}^m)}{\mathcal{D}(\vartheta, \{ p_0, h_0\})} = 1 + \mathcal{O}(p_j-h_j).
}
Combining the results for the three factors we find~\eqref{ff_dressing}.

\section{On the numerical evaluation of the Fredholm determinant}\label{sec:freddet}
The new expression for the thermodynamic form factors \eqref{eq:newD} contains the Fredholm determinant of the generalized resolvent. We restrict here to the single-particle case
\begin{equation}\label{eq:startdet}
 {\rm Det} (\boldsymbol{1} + L^{[p,h]}).
\end{equation}
The numerical evaluation of the kernel $L^{[p,h]}$ is problematic due to the pole in $\lambda' = h$. The most straightforward way is to expand the determinant into traces, using the well known formula
\begin{align}\label{eq:seriesdet}
& {\rm Det} (\boldsymbol{1} + L^{[p,h]}) = \exp \left( \text{Tr} \log   (\boldsymbol{1} + L^{[p,h]}) \right)\nn \\& =  \exp \left(   \text{Tr}  L^{[p,h]} - \frac{1}{2} \text{Tr}  (L^{[p,h]})^2 + \frac{1}{3}\text{Tr}  (L^{[p,h]})^3  + \ldots \right),
\end{align}
with 
\begin{equation}
\text{Tr}  (L^{[p,h]})^n = \int d\lambda_1 \ldots d\lambda_{n-1} L^{[p,h]}(\lambda_1,\lambda_2) \ldots  L^{[p,h]}(\lambda_{n-1},\lambda_1).
\end{equation}
The principal value integrals can be computed using known relations
\begin{equation}
\pint_{-\infty}^{\infty}   {\rm d}\lambda \frac{f(\lambda)}{h - \lambda} =  \text{FT}^{-1}\left( ( i \pi \text{sign} (k) \text{FT}[f] \right)(h),
\end{equation}
with the Fourier transform given by
\begin{equation}
\text{FT} (f) = \int_{-\infty}^{+\infty} {\rm d}x\, e^{i k x} f(x) ,
\end{equation}
and the series \eqref{eq:seriesdet} can be truncated with a small (odd) number of terms. \\

Another approach is to introduce a new kernel
\begin{equation}
 {L}^{[p,h]} (\lambda,\lambda')  = \frac{\tilde{L}^{[p,h]} (\lambda,\lambda') }{ h - \lambda'},
\end{equation}
and decompose $L^{[p,h]}$ as follows 
\begin{equation}
L^{[p,h]}(\lambda,\lambda') = \left[ \frac{\tilde{L}^{[p,h]} (\lambda,\lambda') -  \tilde{L}^{[p,h]} (\lambda,h)}{h-\lambda'}  \right]+ \frac{\tilde{L}^{[p,h]} (\lambda, h) }{h-\lambda'}.
\end{equation}
The first kernel is regular in $\lambda' = h$ and therefore we denote it as $L^{[p,h], \text{reg}}$
\begin{equation}
\frac{\tilde{L}^{[p,h]} (\lambda,\lambda') -  \tilde{L}^{[p,h]} (\lambda,h)}{h-\lambda'} = L^{[p,h], \text{reg}} (\lambda,\lambda') .
\end{equation}
Using that $\frac{\tilde{L}^{[p,h]} (\lambda,h)}{h-\lambda'}$ is a rank one matrix we obtain for the determinant \eqref{eq:startdet}
\begin{equation}
 {\rm Det} (\boldsymbol{1} + L^{[p,h]}) =  {\rm Det}(\boldsymbol{1}+ L^{[p,h],\text{reg}}) \left(1 +  \pint_{-\infty}^{\infty} {\rm d}\lambda {\rm d}\lambda' \frac{\tilde{L}^{[p,h]} (\lambda,h)}{h-\lambda'}  [(1 + L^{[p,h],\text{reg}})^{-1} ] (\lambda ,\lambda') \right).
\end{equation}
We use the usual resolvent technique to compute the inverse of the kernel $(\boldsymbol{1} + L^{[p,h],\text{reg}})^{-1} = \boldsymbol{1}- S^{\text{reg}}$, which gives an integral equation for the kernel $S^{\text{reg}}$
\begin{equation}
S^{\text{reg}}(\lambda,\lambda') +\int {\rm d}\alpha  L^{[p,h],reg}(\lambda,\alpha) S^{\text{reg}}(\alpha,\lambda') = L^{[p,h],\text{reg}}(\lambda,\lambda').
\end{equation}
Finally we obtain 
\begin{align}
 {\rm Det} (\boldsymbol{1} + L^{[p,h]}) & = {\rm Det} (\boldsymbol{1} + L^{[p,h],\text{reg}})  \nn \\& \times  \left(1 +  \pint_{-\infty}^{\infty} {\rm d}\lambda   \frac{\tilde{L}^{[p,h]} (\lambda,h)}{h-\lambda}  - \int {\rm d}\lambda \pint_{-\infty}^{\infty} {\rm d}\lambda'  \frac{\tilde{L}^{[p,h]} (\lambda,h)}{h-\lambda'} S(\lambda,\lambda')\right).
\end{align}
The Fredholm determinant $ {\rm Det} (\boldsymbol{1} + L^{[p,h],\text{reg}})$ can be now easily evaluated with standard numerical techniques~\cite{2010_Bornemann_MC_79}. {  For example by choosing a discretization $\lambda \to \{x_j\}_{j=1}^{N_d}$ and computing the determinant of the finite size matrix $(\delta_{ij} + N_d^{-1} L^{[p,h],\text{reg}}(x_i,x_j))$.}

\end{appendices}

\end{document}